\documentclass[twoside,twocolumn,9pt]{article}
\usepackage{extsizes}
\usepackage[super,sort&compress,comma]{natbib} 
\usepackage[version=3]{mhchem}
\usepackage[left=1.5cm, right=1.5cm, top=1.785cm, bottom=2.0cm]{geometry}
\usepackage{balance}
\usepackage{mathptmx}
\usepackage{sectsty}
\usepackage{graphicx} 
\usepackage{lastpage}
\usepackage[format=plain,justification=justified,singlelinecheck=false,font={stretch=1.125,small,sf},labelfont=bf,labelsep=space]{caption}
\usepackage{float}
\usepackage{fancyhdr}
\usepackage{fnpos}
\usepackage[english]{babel}
\addto{\captionsenglish}{%
  \renewcommand{\refname}{Notes and references}
}
\usepackage{array}
\usepackage{droidsans}
\usepackage{charter}
\usepackage[T1]{fontenc}
\usepackage[usenames,dvipsnames]{xcolor}
\usepackage{setspace}
\usepackage[compact]{titlesec}
\usepackage[%
    bookmarks,%
    bookmarksopen=false,% Klappt die Bookmarks in Acrobat aus
    pdfauthor={Autor},%
    pdftitle={Titel},%
    colorlinks=true,%ture=kein kasten false=kasten
    linkcolor=blue,%red green blue cyan magenta yellow black white
    citecolor=red,%
    urlcolor=cyan,%
]{hyperref}

\usepackage{refcount}
\usepackage{dcolumn}% Align table columns on decimal point
\usepackage{bm}% bold math
\usepackage{braket}
\usepackage{enumitem}
\usepackage{mathptmx}
\usepackage{amssymb} %geschwungene Buchstaben \mathscr 

\usepackage{caption}
\usepackage{subcaption}
\usepackage{siunitx}
\DeclareSIUnit\hartree{\text{E}\ensuremath{_{\text{h}}}}
\DeclareSIUnit\Bau{\text{B}\ensuremath{_0}}
\DeclareSIUnit\angstrom{\text{Å}}

\usepackage{float}
\usepackage{booktabs}
\usepackage{multirow, makecell}
\usepackage{graphicx}

\usepackage{chemfig}
\usepackage{chemexec}
\usepackage{tikz}
\usetikzlibrary{positioning, calc}
\usepackage{pgfplots}
\pgfplotsset{compat=1.17}
\usepackage{diagrams_new}
\usepackage{derivative}
\usepackage[abs]{overpic}
\usepackage{lineno}
\usepackage{cancel}
\definecolor{mycyan}{rgb}{0.380, 0.796, 0.957} % RGB values between 0 and 1

\newcommand{\e}{\operatorname{e}}

\newcommand*{\Hartree}{\text{E}\ensuremath{_{\text{h}}}}

\newcommand{\onlinecite}[1] {\citenum{#1}}

\usepackage{array}
\usepackage{etoolbox}

\makeatletter
               
\def\array@default{c}%

\def\array@row@pre{}%
\def\array@row@pst{}%
\def\tableft@skip{\z@skip}%
\def\tabmid@skip{\z@skip}%\@flushglue
\def\tabright@skip{\z@skip}%
\def\tableftsep{\tabcolsep}%
\def\tabmidsep{\tabcolsep}%
\def\tabrightsep{\tabcolsep}%

\skip\@mpfootins = 0pt
\newenvironment{ruledtabular}{%
\makeatletter
 \def\array@default{v}%
  \def\@halignto{to\hsize}%
  \let\tableft@skip@default\tableft@skip
  \let\tableft@skip\tableft@skip@float
  \let\tabmid@skip@default\tabmid@skip
  \let\tabmid@skip\tabmid@skip@float
  \let\tabright@skip@default\tabright@skip
  \let\tabright@skip\tabright@skip@float
  \let\array@row@pre@default\array@row@pre
  \let\array@row@pre\array@row@pre@float
  \let\array@row@pst@default\array@row@pst
  \let\array@row@pst\array@row@pst@float
}{%
  \let\array@row@pre\array@row@pre@default
  \let\array@row@pst\array@row@pst@default
  \let\tableft@skip\tableft@skip@default
  \let\tabmid@skip\tabmid@skip@default
  \let\tabright@skip\tabright@skip@default
  \let\@halignto\@empty
  \makeatother
}%
\makeatother

\definecolor{cream}{RGB}{222,217,201}

\begin{document}

\pagestyle{fancy}
\thispagestyle{plain}
\fancypagestyle{plain}{
%%%HEADER%%%
\renewcommand{\headrulewidth}{0pt}
}
%%%END OF HEADER%%%

%%%PAGE SETUP - Please do not change any commands within this section%%%
\makeFNbottom
\makeatletter
\renewcommand\LARGE{\@setfontsize\LARGE{15pt}{17}}
\renewcommand\Large{\@setfontsize\Large{12pt}{14}}
\renewcommand\large{\@setfontsize\large{10pt}{12}}
\renewcommand\footnotesize{\@setfontsize\footnotesize{7pt}{10}}
\makeatother

\renewcommand{\thefootnote}{\fnsymbol{footnote}}
\renewcommand\footnoterule{\vspace*{1pt}% 
\color{cream}\hrule width 3.5in height 0.4pt \color{black}\vspace*{5pt}} 
\setcounter{secnumdepth}{5}

\makeatletter 
\renewcommand\@biblabel[1]{#1}            
\renewcommand\@makefntext[1]% 
{\noindent\makebox[0pt][r]{\@thefnmark\,}#1}
\makeatother 
\renewcommand{\figurename}{\small{Fig.}~}
\sectionfont{\sffamily\Large}
\subsectionfont{\normalsize}
\subsubsectionfont{\bf}
\setstretch{1.125} %In particular, please do not alter this line.
\setlength{\skip\footins}{0.8cm}
\setlength{\footnotesep}{0.25cm}
\setlength{\jot}{10pt}
\titlespacing*{\section}{0pt}{4pt}{4pt}
\titlespacing*{\subsection}{0pt}{15pt}{1pt}
%%%END OF PAGE SETUP%%%

%%%FOOTER%%%
\fancyfoot{}
\fancyfoot[LO,RE]{\vspace{-7.1pt}\includegraphics[height=9pt]{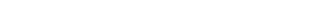}}
\fancyfoot[CO]{\vspace{-7.1pt}\hspace{11.9cm}\includegraphics{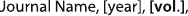}}
\fancyfoot[CE]{\vspace{-7.2pt}\hspace{-13.2cm}\includegraphics{head_foot/RF}}
\fancyfoot[RO]{\footnotesize{\sffamily{1--\pageref{LastPage} ~\textbar  \hspace{2pt}\thepage}}}
\fancyfoot[LE]{\footnotesize{\sffamily{\thepage~\textbar\hspace{4.65cm} 1--\pageref{LastPage}}}}
\fancyhead{}
\renewcommand{\headrulewidth}{0pt} 
\renewcommand{\footrulewidth}{0pt}
\setlength{\arrayrulewidth}{1pt}
\setlength{\columnsep}{6.5mm}
\setlength\bibsep{1pt}
%%%END OF FOOTER%%%

%%%FIGURE SETUP - please do not change any commands within this section%%%
\makeatletter 
\newlength{\figrulesep} 
\setlength{\figrulesep}{0.5\textfloatsep} 

\newcommand{\topfigrule}{\vspace*{-1pt}% 
\noindent{\color{cream}\rule[-\figrulesep]{\columnwidth}{1.5pt}} }

\newcommand{\botfigrule}{\vspace*{-2pt}% 
\noindent{\color{cream}\rule[\figrulesep]{\columnwidth}{1.5pt}} }

\newcommand{\dblfigrule}{\vspace*{-1pt}% 
\noindent{\color{cream}\rule[-\figrulesep]{\textwidth}{1.5pt}} }

\makeatother
%%%END OF FIGURE SETUP%%%

%%%TITLE, AUTHORS AND ABSTRACT%%%
\twocolumn[
  \begin{@twocolumnfalse}
%{\includegraphics[height=30pt]%{head_foot/PCCP}\hfill\raisebox{0pt}[0pt][0pt]{\includegraphics[height=55pt]{head_foot/RSC_LOGO_CMYK}}\\[1ex]
%\includegraphics[width=18.5cm]{head_foot/header_bar}}\par
\vspace{1em}
\sffamily
\begin{tabular}{m{4.5cm} p{13.5cm} }

\includegraphics{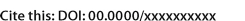} & \noindent\LARGE{\textbf{Finite-Field Cholesky Decomposed Coupled-Cluster Techniques (ff-{CD-CC}): Theory and Application to Pressure Broadening of Mg by a He Atmosphere and a Strong Magnetic Field$^\dag$}} \\%Article title goes here instead of the text "This is the title"
\vspace{0.3cm} & \vspace{0.3cm} \\

 & \noindent\large{Simon Blaschke\textit{$^{a,b}$}, Marios-Petros Kitsaras\textit{$^{b}$} and Stella Stopkowicz$^{\ast}$\textit{$^{b,c}$}} \\%Author names go here instead of "Full name", etc.

\includegraphics{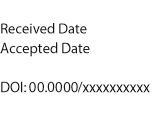} & \noindent\normalsize{For the interpretation of spectra of magnetic stellar objects such as magnetic white dwarfs (WDs) highly accurate quantum chemical predictions for atoms and molecules in finite magnetic field are required. 
Especially the accurate description of electronically excited states and their properties requires established methods such as those from coupled-cluster (CC) theory. 
However, respective calculations are computationally challenging even for medium-sized systems. 
Cholesky decomposition (CD) techniques may be used to alleviate memory bottlenecks. 
In finite-field computations, the latter are increased due to the reduction of permutational symmetry within the electron-repulsion-integrals (ERIs) as well as the need for complex-valued data types. 
CD enables a memory-efficient, approximate description of the ERIs with rigorous error control and thus the treatment of larger systems at the CC level becomes feasible. 
In order to treat molecules in a finite magnetic field (ff), we present in this work the working equations of the left and right-hand side equations for ff-EOM-CD-CCSD for various EOM flavours as well as for the approximate ff-EOM-CD-CC2 method. 
The methods are applied to the study of the modification of the spectral lines of a magnesium transition by a helium atmosphere that can be found in magnetic WD stars.} \\%The abstrast goes here instead of the text "The abstract should be..."

\end{tabular}

 \end{@twocolumnfalse} \vspace{0.6cm}

  ]
%%%END OF TITLE, AUTHORS AND ABSTRACT%%%

%%%FONT SETUP - please do not change any commands within this section
\renewcommand*\rmdefault{bch}\normalfont\upshape
\rmfamily
\section*{}
\vspace{-1cm}

%%%FOOTNOTES%%%

\footnotetext{\textit{$^{a}$~Department Chemie, Johannes Gutenberg-Universität Mainz, Duesbergweg 10-14, D-55128 Mainz, Germany.}}
\footnotetext{\textit{$^{b}$~Fachrichtung Chemie, Universit{\"a}t des Saarlandes, Campus B2.2, D-66123 Saarbr{\"u}cken, Germany.} E-mail: stella.stopkowicz@uni-saarland.de}
\footnotetext{\textit{$^{c}$~Hylleraas Centre for Quantum Molecular Sciences, Department of Chemistry, University of Oslo, P.O. Box 1033, N-0315 Oslo, Norway.}}

%Please use \dag to cite the ESI in the main text of the article.
%If you article does not have ESI please remove the the \dag symbol from the title and the footnotetext below.
\footnotetext{\dag~Supplementary Information available}
%additional addresses can be cited as above using the lower-case letters, c, d, e... If all authors are from the same address, no letter is required

%%%END OF FOOTNOTES%%%

\expandafter\let\expandafter\oldtabulars\csname tabular\endcsname
\expandafter\let\expandafter\endoldtabulars\csname endtabular\endcsname

\RenewDocumentEnvironment{tabular}{ O{c} m }{%
  \begin{oldtabulars}[#1]{#2}
    \midrule\midrule    
}{%
  \\[-10pt]\midrule\midrule
 \end{oldtabulars}%
}

\expandafter\let\expandafter\oldtabularsstar\csname tabular*\endcsname
\expandafter\let\expandafter\endoldtabularsstar\csname endtabular*\endcsname

\RenewDocumentEnvironment{tabular*}{ m O{c} m }{%
  \begin{oldtabularsstar}{#1}[#2]{#3}
  \midrule\midrule  
}{%
  \\[-10pt]\midrule\midrule
 \end{oldtabularsstar}%
}

%%%MAIN TEXT%%%%

%%%%%%%%%%%%%%%%%%%%%%%%%%%%%%%%%%%%%%
\section{Introduction}\label{sec:introduction}
%%%%%%%%%%%%%%%%%%%%%%%%%%%%%%%%%%%%%%
%
In the last decades, there has been a growing interest among astrophysicists in the study of (magnetic) white dwarfs (WDs) as part of the exploration of stellar evolution.\cite{Kemp1970,Schmelcher.AAp.1998,Schmelcher.J.Phys.B:At.Mol.Opt.Phys.1999,Wickramasinghe.Publ.Astron.Soc.Pac.2000,Kepler2013,Hollands2023} Besides neutron stars and black holes, WDs represent the by far most common endpoint in the life cycle of stars, i.e., 95\% of all stars become WDs. 
Of those approximately 25\% exhibit magnetic properties, with magnetic field strengths reaching up to \SI{0.4}{\Bau} ($\SI{1}{\Bau}=\SI{235052}{\tesla}$), where electronic and magnetic forces are at the same order of magnitude. In order to assign spectra from strongly magnetized WDs, predictions for transition wavelengths and intensities are required. Two main considerations have to be taken into account: First, the magnetic field has to be accounted for in the Hamiltonian in an explicit manner. It cannot be treated perturbatively as done in the weak-field limit for e.g. nuclear magnetic resonance (NMR), electron paramagnetic resonance (EPR) or magnetic circular dichroism (MCD) spectroscopy.\cite{Gauss.Molecular.Properties.2000}
Second, as electronic structure, bonding mechanisms, excitation energies and properties change significantly under the influence of such strong fields and no laboratory experiments are available, accurate computational predictions are crucial.\newline
Coupled-cluster (CC) methods\cite{Shavitt2009.MBPT,Crawford2000} allow for a precise electron-correlation treatment. However, respective CC calculations are highly demanding due to the steep scaling and memory bottlenecks.\newline
In the absence of external magnetic fields, significant efforts have been made to tackle these issues and to improve the computational cost for standard CC methods. For example, Pulay and S{\ae}b{\o}\cite{Pulay.Chem.Phys.Lett.1985,Saebo.Annu.Rev.Phys.Chem.1993} introduced an approach which exploits the local nature of the short-range dynamic electron correlation that ultimately leads to sparse tensors, i.e., local correlation approaches. This ansatz uses orbital localization\cite{Boys.Rev.Mod.Phys.1960, FosterBoys.Rev.Mod.Phys.1960, EdmistonRuedenberg.Rev.Mod.Phys.1963, PipekMezey.J.Chem.Phys.1989} for the occupied space and only a subset of excitations to the virtual space within spatial vicinity is treated. Using the projected atomic orbital (PAO)\cite{Sinanoglu.Adv.Chem.Phys.1964, Nesbet.Adv.Chem.Phys.1965, Pulay.Chem.Phys.Lett.1983} representations for the virtual space efficient local CC implementations haven been realized by Werner and co-workers at the CC and Equation-of-Motion (EOM)-CC levels (CCSD \cite{Werner.J.Chem.Phys.1996,Schuetz.J.Chem.Phys.2001,Werner.J.Chem.Phys.2003}, CCSD(T)\cite{Schuetz.Phys.Chem.Chem.Phys.2002}, and CC2\cite{Schuetz.J.Chem.Phys.2006}), respectively. In addition, Neese et al.\cite{Neese.J.Chem.Phys.2009a} revived the use of the pair natural orbital (PNO)\cite{Loewdin.Phys.Rev.1955, Krauss.J.Chem.Phys.1966, Meyer.Int.J.Quantum.Chem.1971} representation for the virtual space which led to the highly successful domain based (D)LPNO-CCSD\cite{Neese.J.Chem.Phys.2009, Neese.J.Chem.Phys.2013, Neese.J.Chem.Phys.2016} scheme. 
An extension of this ansatz to excited states has been described by Dutta et al.\cite{Dutta2016,Dutta2018,Dutta2019}\newline
Other approaches to lower the computational bottleneck are fragmentation-based schemes such as fragment molecular orbital (FMO)\cite{Kitaura.Chem.Phys.Lett.1999, Fedorov.J.Chem.Phys.2005, Kitaura.J.Phys.Chem.A.2007} and molecular fractionation with conjugate caps (MFCC)\cite{Zhang.J.Chem.Phys.2003}. Here, the whole system is divided into fragments and treated via a many-body expansion. The fragmentation methods are mainly used for the description of large clusters of weakly interacting monomers or the description of large biochemical molecules like peptides. These are molecular systems which can often also be addressed by the related multilevel embedding approaches, like QM/MM\cite{Warshel.J.Am.Chem.Soc.1972,Warshel.J.Mol.Biol.1976} or QM/QM\cite{Svensson.J.Phys.Chem.1996} which allow for the accurate treatment of a desired subsystem while still describing the polarization and dynamics of the surrounding medium.\newline
Instead of exploiting the sparsity of tensors via spatial locality of orbital subspaces or molecular fragments another ansatz is to use a rank-reduced approximation of the electron-repulsion integrals (ERIs). This ansatz is applied in density fitting (DF) also termed resolution-of-identity approximation (RI)\cite{Whitten.1973, Baerends1973, Dunlap1979, Vahtras.Chem.Phys.Lett.1993, Feyereisen.Chem.Phys.Lett.1993, Komornicki.J.Chem.Phys.1993, Dunlap.I.2000, Dunlap.II.2000, Dunlap.III.2000, Dunlap.IV.2003} or Cholesky decomposition (CD)\cite{LinderbergJ.Quantum.Chem.1977, KochJ.Chem.Phys.2003, KochJ.Chem.Phys.2019, Zhang.J.Phys.Chem.A.2021, Blaschke.J.Chem.Phys.2022, Gauss.Mol.Phys.2022} approaches. 
Using either approach the ERIs can represented by the product of rank three tensors. 
The RI approach uses a pre-optimised\cite{Eichkorn1995, Eichkorn1997, Weigend1998, Weigend2006, Weigend2002, Weigend2007} or automatically generated auxiliary basis\cite{Pedersen.J.Chem.Phys.2007, Aquilante.J.Chem.Phys.2009, NeeseJ.Chem.TheoryComput.2017, Lehtola.2021, Lehtola2023} in order to approximate the product density in the ERIs, whereas for the CD approach a CD of the ERIs is performed. Defining the Cholesky basis (CB) as collection of product densities corresponding to the pivots of the selected Cholesky vectors \textcolor{red}{(CV)}, the latter two approaches become equivalent if the Cholesky basis is chosen as the auxiliary basis for RI.\cite{Pedersen.Theor.Chem.Acc.2009,LinearScalingTechniques}
Since the reintroduction of CD by Koch, S{\'{a}}nchez de Mer{\'{a}}s and Pedersen\cite{KochJ.Chem.Phys.2003} in 2003, many methods exploiting the CD representation of the ERIs have been developed.\cite{Aquilante.Chem.Phys.Lett.2007, Koch.MCSCF.J.Chem.Phys.2008, Aquilante.2008, Pedersen.J.Chem.Phys.2008, Pedersen.Int.J.Quantum.Chem.2013, Pedersen.J.Chem.Phys.2015, KochJ.Chem.Phys.2019, Zhang.J.Phys.Chem.A.2021, Burger.2021, Blaschke.J.Chem.Phys.2022, Gauss.Mol.Phys.2022, Nottoli.Mol.Phys.2021, Nottoli.J.Chem.Theor.Comput.2021, Bozkaya.JCTC.2022, Nottoli2022} The developments in CC theory include \textcolor{red}{CD-}(EOM)-CCSD\cite{Krylov.2013, GaussJ.Chem.Phys.2019, Nottoli2023, Zhang2024, Uhlirova2024}, approximate CD-CC2 linear response\cite{Pedersen.J.Chem.Phys.2004} as well as analytic gradients\cite{GaussJ.Chem.Phys.2019,Koch.J.Chem.Phys.2022}.
While RI and CD achieve memory savings due to the rank reduction of the ERI tensor, the computational savings are mostly due to a favourable prefactor and reduced I/O overhead.
However, the overall scaling is generally not changed. Further factorization of the ERIs (or their approximate representations) yields a method that decouples all indices of the ERI and thus also handles exchange-like terms well, as, for example, tensor hypercontraction (THC) implementations of (EOM)-CC2\cite{Hohenstein.J.Chem.Phys.2013,Hohenstein.J.Phys.Chem.2013} and (EOM)-CCSD\cite{MartinezJ.Chem.Phys.III.2012,MartinezJ.Chem.Phys.2014,Datar2024}.\newline
In addition to the decomposition of the ERI the decomposition of the wave-function amplitudes which can be employed concurrently, was explored as well. This ansatz is conceptually related to earlier work on the Laplace transformation of the orbital denominator by Häser and Almlöf.\cite{Almloef.Chem.Phys.Lett.1991,Haeser.J.Chem.Phys.1992} Later, Aquilante et al.\cite{Aquilante.Chem.Phys.Lett.2007} recognized that the MP2 amplitudes, as a negative definite matrix, can be Cholesky decomposed which was combined with Cholesky decomposed ERIs in the context of quintic scaling SOS-MP2. Equivalently Koch and {S{\'{a}}nchez de Mer{\'{a}}s\cite{Koch.J.Chem.Phys.2000,Koch.Int.J.Quantum.Chem.2010} utilized a CD of the orbital denominator in the perturbative treatment of the triples contribution within CCSD(T). Kinoshita, Hino and Bartlett\cite{Bartlett.J.Chem.Phys.2003,Bartlett.J.Chem.Phys.2004} proposed the decomposition of the CC amplitudes. As the CC amplitudes are no longer negative definite, they applied an alternative decomposition method, i.e., singular value decomposition (SVD). Following this scheme, various rank reduced (RR) CC schemes using orthogonal projectors were explored.\cite{Parrish.J.Chem.Phys.2019,Hohenstein.J.Chem.Phys.2019,Hohenstein.J.Chem.Phys.2022,Lesiuk.J.Comput.Chem.2019,Lesiuk.J.Chem.Phys.2022,Jiang.J.Chem.Theory.Comput.2023}\newline
Going from the field-free to the finite magnetic field treatment is associated with additional computational cost due to the complex nature of the wave function. 
This necessitates the use of complex arithmetic in electronic-structure methods which usually implies the need of new software implementations. For such implementations the memory requirements double and floating point operations (FLOPs) quadruple\footnote[3]{The requirements triple with the use of optimised BLAS library for matrix multiplication \textsc{zgemm3m}.} compared to real floating point arithmetic. Additionally, a reduction of permutational symmetry of the ERIs from eightfold to fourfold results in an even greater increase in costs. Furthermore, in most cases, the symmetry of the system is lowered due to the presence of the magnetic field. Lastly, large uncontracted basis sets are required to adequately describe the anisotropy induced by the magnetic field.\cite{Stopkowicz.2015,Astrom2023} This anisotropy is generally not well described by standard isotropic Gaussian basis functions.\cite{Aldrich1979,Schmelcher1988,Zhu2017} Consequently, incorporating an external magnetic field results in a substantial increase in computational cost.\newline
In recent years, efficient implementations of various finite-field~(ff) methods have been realized.\cite{Reynolds.Phys.Chem.Chem.Phys.2015, ReynoldsJ.Chem.Phys.2018, pausch.klopperMolPhys2020, Pausch.Front.Chem.2021, TealeJ.Chem.TheoryComput.2021, Pausch.J.Phys.Chem.Lett.2022, Pausch.J.Chem.TheoryComput.2022, Blaschke.J.Chem.Phys.2022, Teale.J.Chem.Theory.Comput.2022} 
Though, at the present date efficient implementations of ff-CC using any or a combination of aforementioned methods are scarce. To the best of our knowledge solely an implementation of RI-CC2\cite{Pausch.Front.Chem.2021} and the embedded fragment method~(EFM) by Speake et al.\cite{Teale.J.Chem.Theory.Comput.2022} are available.\newline
Therefore, there is a need for an adaption of the CCSD method with the following features: The accuracy of the adaptation has to be controllable. Further, it has to reduce the memory requirements so that medium to large systems can be described, despite the steep scaling of $\mathcal{O}(N^6)$ while keeping computational cost as low as possible. Thus, for our implementation we chose the CD which provides rigorous error control via a single threshold and low memory requirements for the handling of the CVs in comparison to the full ERI. Apart from the well defined error control, within ff calculations, a further advantage of CD over RI/DF approaches is the flexibility of the on-the-fly determined Cholesky basis to adapt to changes in the ERI tensor due to the magnetic field via its strength and relative orientation with respect to the molecule.\cite{Blaschke.J.Chem.Phys.2022} 
RI auxiliary basis sets, on the other hand, have so far only been optimized in the field-free case and their an on-the-fly generation\cite{Pedersen.J.Chem.Phys.2007, Aquilante.J.Chem.Phys.2009, NeeseJ.Chem.TheoryComput.2017, Lehtola.2021,Lehtola2023} for use in a magnetic field is currently not available. 
However, one might argue that RI is computationally more efficient as the auxiliary basis is usually chosen to contain only one-center functions. 
Thus, RI requires the evaluation of at most three-center integrals and for ff applications, auxiliary basis sets can be selected as real, retaining the permutational symmetry and thus memory efficiency.\cite{Reynolds.Phys.Chem.Chem.Phys.2015,pausch.klopperMolPhys2020} 
The CD scheme on the other hand necessitates the evaluation of four-center-integrals due to the fact that, unlike the RI basis, the Cholesky basis is generally not centered on one atom. 
Still, its use may be preferable as the anisotropy of the magnetic field demands a high flexibility of the basis set and it is still an open question how well these effects are captured by a one-center approximation. 
In addition, while the additional memory requirements for CD vs. RI are valid concern for methods in which the integral evaluation is a bottleneck, this is much less relevant in the case for CC theory. Thus the rigorous error control and the flexibility to include magnetic field effects in the Cholesky basis makes CD our preferred choice.\newline
In section~\ref{sec:theory} a brief review of the CD of the ff-ERIs is presented. The decomposition of ERIs over London orbitals and application using M\o ller-Plesset perturbation theory was already presented in Ref.~\onlinecite{Blaschke.J.Chem.Phys.2022}. In the following section~\ref{sec:CoupledCluster} we present the working equations of the left and right-hand side equations of different variants of CD-EOM-CCSD as well as the approximate CD-EOM-CC2 method. Section~\ref{sec:implementation} discusses the details of the implementation while in section~\ref{sec:Validation} the validation is carried out on a small test set of molecules monitoring the error for various Cholesky thresholds. Using the new developments we investigate the spectral lines of magnesium atoms in dense helium atmospheres on magnetic stellar objects such as magnetic WDs in section~\ref{sec:Application}. 
We start by studying the pressure broadening of transition lines in the field-free case as well as a finite field on the MgHe dimer as model system confirming the observed (asymmetric) blue-shifted band shape and unraveling a strong dependence on the orientation with respect to the external magnetic field. 
For helium atmospheres which exhibit large densities, the MgHe dimer becomes an insufficient model, as interperturber interactions become more probable. In a first-order approximation, we transition to an explicit description of the surrounding atmosphere by treating \ce{MgHe12} and \ce{MgHe56} clusters in a hcp structure, which are employed to describe the transition wavelength under such conditions. For these clusters the use of CD becomes mandatory as the memory requirements otherwise become prohibitively large.

%%%%%%%%%%%%%%%%%%%%%%%%%%%%%%%%%%%%%%%%%
\section{Theory}\label{sec:theory}
%%%%%%%%%%%%%%%%%%%%%%%%%%%%%%%%%%%%%%%%%
%%%%%%%%%%%%%%%%%%%%%%%%%%%%%%%%%%%%%%%%%
\subsection{Finite-Field Hamiltonian}\label{sec:Finite_Field_Hamiltonian}
%%%%%%%%%%%%%%%%%%%%%%%%%%%%%%%%%%%%%%%%%

In order to describe the electronic structure of a molecule in a strong magnetic field, the Lorentz forces have to be treated in a non-perturbative manner. 
The molecular electronic finite-field Hamiltonian in a uniform external magnetic field in $z$-direction is given in atomic units as
\begin{equation}\label{eq:Hamiltonoperator_im_magnetfeld}
    \hat{H}_{\text{el}}=\hat{H}_0+\frac{1}{2}B_z\cdot \hat{L}_{z}^{\text{O}}+B_z\cdot \hat{S}_{z}+\frac{1}{8}\sum_{i}^{N}B_z^2\left({x^{\text{O}}_i}^2+{y^{\text{O}}_i}^2\right)\;.
\end{equation}
Here $\hat{H}_0$ is the field-free non-relativistic electronic Hamiltonian in the framework of the screened Born-Oppenheimer approximation.\cite{Schmelcher1988} The remaining terms depend on the magnetic field $B_z$. The paramagnetic terms are linear in the magnetic field and scale with the total spin $\hat{S}_{z}$ and the canonical angular momentum operator $\hat{\boldsymbol{L}}^{\text{O}}=-\sum_{i}^{N}\text{i}\boldsymbol{r}_i^{\text{O}}\times \boldsymbol{\nabla}_i$, respectively. $\hat{L}^{\text{O}}_{z}$ depends on the distance vector $\boldsymbol{r}_i^{\text{O}}$ of the electron $i$ relative to the arbitrary gauge origin $\boldsymbol{O}$. 
The diamagnetic term is quadratic in $B_z$ which constitutes a confining potential for the directions perpendicular to the magnetic field axis.\newline
The angular momentum operator $ \hat{L}_{z}^{\text{O}}$ and the coordinates ${x^{\text{O}}_i}$ and ${y^{\text{O}}_i}$ depend on the location of the gauge origin $\boldsymbol{O}$. 
As approximated wave functions do not show correct transformation behaviour with respect to a gauge origin transformation observables are dependent on the choice of the gauge origin. This is solved by using gauge including atomic orbitals (GIAOs) also termed London orbitals\cite{London,TellgrenJ.Chem.Phys.2008}
\begin{equation}
    \omega_{\mu}=\text{e}^{-i\boldsymbol{k}\boldsymbol{r}}\chi_{\mu}
    \label{eq:LondonAO}
\end{equation}
which ensure gauge-origin invariance. They are constructed by a product of a standard Gaussian $\chi_\mu$ centered at $\boldsymbol{K}_\mu$ with a complex phase factor in which $\boldsymbol{k}=\frac{1}{2}\boldsymbol{B}\times (\boldsymbol{K}_\mu-\boldsymbol{O})$. The evaluation of integrals over GIAOs is discussed in detail in Refs.~\onlinecite{TellgrenJ.Chem.Phys.2008, Teale.J.Chem.TheoryComput.2017, Blaschke.J.Chem.Phys.2024}.

%%%%%%%%%%%%%%%%%%%%%%%%%%%%%%%%%%%%%%
\subsection{Cholesky Decomposition of ERIs over London Orbitals}\label{sec:CDoverGIAOs}
%%%%%%%%%%%%%%%%%%%%%%%%%%%%%%%%%%%%%%

The matrix $\bm{V}$ of the ERIs over London orbitals with elements
\begin{equation}
    \left(\mu\nu|\sigma\rho\right)=\int\int\omega_{\mu}^*(\boldsymbol{r}_1)\omega_{\nu}(\boldsymbol{r}_1)r_{12}^{-1}\omega_{\sigma}^*(\boldsymbol{r}_2)\omega_{\rho}(\boldsymbol{r}_2)\text{d}\boldsymbol{r}_1\text{d}\boldsymbol{r}_2\;,
\end{equation}
is semi positive definite and may thus be factorized as $\bm{V} =\sum_J\bm{\ell}^J\bm{\ell}^{J\dagger}$ using the CD scheme.
The elements $L_{\sigma\rho}^J$ of the CV $\bm{\ell}^J$ are determined iteratively by a partial pivoting procedure\cite{KochJ.Chem.Phys.2003}. The pivot corresponds to the largest element of the updated diagonal of the two-electron integral matrix $D^{J}_{\mu\nu}$ which is given as
\begin{equation}
    D^{J}_{\mu\nu}=\left(\mu\nu|\nu\mu\right)-\sum_{K=1}^{J-1}L^{K}_{\mu\nu}L^{K*}_{\mu\nu}\;.
\end{equation}
As in an external magnetic field the equivalence $(\mu\nu|\nu\mu)=(\mu\nu|\mu\nu)$ does not hold, it is crucial to consider that the diagonal of the ERIs is given as $(\mu\nu|\nu\mu)$ to ensure that the ERI is positive semi-definite.
The integral column $\left(\sigma\rho|\mu\nu\right)$ corresponding to the pivot element is updated by the contributions of previously determined CVs and finally normalized by the updated diagonal element to yield the elements of the new CV
\begin{equation}
    L^{J}_{\sigma\rho}=\frac{1}{\sqrt{D^{J}_{\mu\nu}}}\left[\left(\sigma\rho|\mu\nu\right)-\sum_{K=1}^{J-1}L^{K}_{\sigma\rho}L^{K*}_{\nu\mu}\right]\;.
    \label{eq:cholvec}
\end{equation}
Truncating the decomposition at the iteration where all remaining updated diagonal elements are smaller than a chosen threshold $\tau=10^{-\delta}$, where $\delta$ is the Cholesky parameter, the ERI is then approximated by
\begin{equation}
     \begin{split}\label{eq:Cholesky_Zweielektronenmatrix_AO_Basis}
        \left(\mu\nu|\sigma\rho\right)\approx\sum_{J=1}^{N_{\text{CH}}}L_{\mu\nu}^{J}L_{\rho\sigma}^{J*}\;.
    \end{split}
\end{equation}
The number of required vectors $N_{\text{CH}}$ is significantly smaller than the number of vectors for the full decomposition since only linear independent Cholesky basis functions (within the limits of the Cholesky threshold) enter the vector manifold. The consequences for the decomposition algorithm and the influence of finite fields on the structure of the ERI were discussed in Refs.~\onlinecite{Blaschke.J.Chem.Phys.2022} and \onlinecite{Gauss.Mol.Phys.2022}.\newline
Alternatively a two-step procedure\cite{KochJ.Chem.Phys.2019, Zhang.J.Phys.Chem.A.2021} can be employed to evaluate the Cholesky vectors. Here in a first step the Cholesky basis which contains the elements of the product densities that are chosen as pivots, is determined equivalently to the aforementioned pivoting procedure. By restricting the decomposition to the potential pivot indices, Folkestad et al.\cite{KochJ.Chem.Phys.2019} showed that the determination of the Cholesky basis can be done even more efficiently. In addition, bookkeeping of the already calculated integrals can prevent repeated calculation as shown by Zhang et al.\cite{Zhang.J.Phys.Chem.A.2021} In a second step the Cholesky vectors are determined. Therefore, the equality of RI and CD is invoked when using the Cholesky basis $|K)$ as auxiliary basis. The integrals can thus be expressed as
\begin{equation}\label{eq:RI_expression_for_CD}
    \left(\mu\nu|\sigma\rho\right)=\sum_{KL}\left(\mu\nu|K\right)\left(K|L\right)^{-1}\left(L|\sigma\rho\right)
\end{equation}
with
\begin{align}
    \left(\mu\nu|K\right) &= \iint\omega_{\mu}^*(\boldsymbol{r}_1)\omega_{\nu}(\boldsymbol{r}_1)r_{12}^{-1}K(\boldsymbol{r}_2)\;\text{d}\boldsymbol{r}_1\text{d}\boldsymbol{r}_2\;,\\
    \left(K|L\right) &= \iint K^*(\boldsymbol{r}_1)r_{12}^{-1}L(\boldsymbol{r}_2)\;\text{d}\boldsymbol{r}_1\text{d}\boldsymbol{r}_2\;,\\
    \left(L|\sigma\rho\right) &= \iint L^*(\boldsymbol{r}_1)r_{12}^{-1}\omega_{\sigma}^*(\boldsymbol{r}_2)\omega_{\rho}(\boldsymbol{r}_2)\;\text{d}\boldsymbol{r}_1\text{d}\boldsymbol{r}_2\;.
\end{align}
The overlap of the Cholesky basis $S_{KL}=\left(K|L\right)$ is positive-semi definite\footnote[4]{The Cholesky basis $|K)$ refers to the product densities $|\mu\nu)$, the ket $(K|$ is defined as $(\nu\mu|$ resulting in a Hermitian matrix. As such, the diagonal is defined as $(K|K)=(\mu\nu|\nu\mu)$.} and may be Cholesky decomposed
\begin{equation}
    \mathbf{S}=\mathbf{M}\mathbf{M}^\dagger\;.
\end{equation}
With the Cholesky representation of the overlap Eq.~\ref{eq:RI_expression_for_CD} results in
\begin{equation}
\begin{split}
        \left(\mu\nu|\sigma\rho\right)&=\sum_{KL}\left(\mu\nu|K\right)(\mathbf{M}\mathbf{M}^\dagger)^{-1}_{KL}\left(L|\sigma\rho\right)\\
        &=\sum_{KL}\left(\mu\nu|K\right)(\mathbf{M}^{-\dagger}\mathbf{M}^{-1})_{KL}\left(L|\sigma\rho\right)\\
        &=\sum_{J}\biggl[\sum_K\left(\mu\nu|K\right)(\mathbf{M}^{-\dagger})_{KJ}\biggr]\biggl[\sum_{L}(\mathbf{M}^{-1})_{JL}\left(L|\sigma\rho\right)\biggr]\\
\end{split}
\end{equation}
using the identity $(\mathbf{A}\mathbf{B})^{-1}=\mathbf{B}^{-1}\mathbf{A}^{-1}$ for square matrices $\mathbf{A}$ and $\mathbf{B}$ and $\mathbf{M}^{-\dagger}=(\mathbf{M}^\dagger)^{-1}$. Comparison with Eq.~\ref{eq:Cholesky_Zweielektronenmatrix_AO_Basis} yields the equation for determining the Cholesky vectors:
\begin{equation}\label{eq:Bestimmungsgl_CV_2Step}
    L_{\mu\nu}^J = \sum_K\left(\mu\nu|K\right)(\mathbf{M}^{-\dagger})_{KJ}\;.
\end{equation}
Evaluating Eq.~\ref{eq:Bestimmungsgl_CV_2Step} can be done using efficient BLAS and LAPACK routines resulting in a superior efficiency compared the the pivoting procedure.\newline
For correlated calculations a sequential transformation of the CVs into the basis of molecular orbitals (MOs)
\begin{equation}\label{eq:transformierte_Cholesky_Vektoren}
    L_{pq}^{J}=\sum_{\mu\nu}C_{\mu p}^{*}L_{\mu \nu}^{J}C_{\nu q}
\end{equation}
is performed. The corresponding MO integrals are expressed analogously as
\begin{equation}\label{eq:Cholesky_Zweielektronenmatrix_MO_Basis}
    \left(pq|rs\right)\approx \sum_{J}^{N_{\text{CH}}}L_{pq}^{J}L_{sr}^{J*}
\end{equation}
and may readily be used in a subsequent CC treatment. 

%%%%%%%%%%%%%%%%%%%%%%%%%%%%%%%%%%%%%%
\subsection{Cholesky Decomposition in finite-field CC theory}\label{sec:CoupledCluster}
%%%%%%%%%%%%%%%%%%%%%%%%%%%%%%%%%%%%%%

In CC theory\cite{Cizek1966, Cizek1969, Bartlett1978, Pruvis.Bartlett.J.Chem.Phys.1982, Stanton1991, Bartlett.2007, Stopkowicz.2015, Stopkowicz.2018, Shavitt2009.MBPT, Crawford2000}, an exponential approach is chosen for the wave function
\begin{equation}\label{eq:CC-Wellenfunktion}
\begin{split}
    \ket{\Psi_{\text{CC}}}&=\e^{\hat{T}}\ket{\Phi_0}\\
    &=\left(1+\hat{T}+\frac{\hat{T}^2}{2!}+\frac{\hat{T}^3}{3!}+\hdots\right)\ket{\Phi_0}\;.
    \end{split}
\end{equation}
The exponential term can be represented by a series expansion as a sum of cluster operators $\hat{T}$ of different powers. The operators act on a reference determinant $\ket{\Phi_0}$, for which the HF wave function is usually chosen. The cluster operator $\hat{T}$ of an $N$-electron system is obtained as the sum of $N$ excitation operators~$\hat{T}_n$:
\begin{equation}\label{eq:Clusteroperator}
    \begin{split}
        \hat{T}&=\hat{T}_1+\hat{T}_2+\hdots+\hat{T}_N\\
            &=\sum_{n=1}^{N}\left(\frac{1}{n!}\right)^2\sum_{ij\hdots ab}t_{ij\hdots}^{ab\hdots}\hat{a}_{a}^{\dagger}\hat{a}_{b}^{\dagger}\hdots\hat{a}_{j}\hat{a}_{i}\\
            &=\sum_I t_I \hat{\mu}_I\;.
    \end{split}
\end{equation}
The excitation operators consist of the weighting coefficients, the amplitudes $t_{I}$, and a string of quasiparticle creation operators $\hat{\mu}_I$, i.e., $\hat{a}_{a}^{\dagger}$ and $\hat{a}_{i}$. The indices $a,b,c\hdots$ as well as  $i,j,k\hdots$ denote virtual and occupied orbitals, respectively. 
The CC energy and amplitude equations are given by
\begin{equation}\label{eq:CC-Energiegleichung}
    \bra{\Phi_0}\bar{H}\ket{\Phi_0}=E_{\text{CC}}\;,
\end{equation}
\begin{equation}\label{eq:CC-Amplitudengleichung}
    \Omega_I(\mathbf{t})=\bra{\Phi_{I}}\bar{H}\ket{\Phi_0}=0\;
\end{equation}
with $\bar{H}=\e^{-\hat{T}}\hat{H}\e^{\hat{T}}$ as the similarity-transformed Hamiltonian.
Restriction of the cluster operator to contain only single and double excitations as well as restricting the projection space for the amplitude to the subspace of excited determinants $\bra{\Phi_{I}}$ obtained by the action of the cluster operator on the reference determinant leads to the CCSD truncation scheme.\cite{Pruvis.Bartlett.J.Chem.Phys.1982,Stanton1991,Bartlett.2007}\newline
Within the CC2\cite{Christiansen.Chem.Phys.Lett.1995, Christiansen.J.Chem.Phys.1995, Koch.J.Chem.Phys.1997, Hattig.J.Chem.Phys.2000, Kitsaras.J.Chem.Phys.2024} framework an approximate CCSD formulation is achieved.
By introducing the $T_1$-transformed Hamiltonian $\tilde{H}=\e^{-\hat{T_1}}\hat{H}\e^{\hat{T_1}}$ the CCSD amplitude equations can be rewritten as
\begin{equation}
    \bra{\Phi_{i}^{a}}\tilde{H}+\left[\tilde{H},\hat{T}_2\right]\ket{\Phi_0}=0\;,
\end{equation}
\begin{equation}
    \bra{\Phi_{ij}^{ab}}\tilde{H}+\left[\tilde{H},\hat{T}_2\right]+\frac{1}{2}\left[\left[\tilde{H},\hat{T}_2\right],\hat{T}_2\right]\ket{\Phi_0}=0\;.
\end{equation}
A M\o ller-Plesset type partitioning of the Hamiltonian $\hat{H}=\hat{F}+\hat{U}$ is performed and only contributions through first order in perturbation theory are considered. 
The singles contributions are treated as zeroth order parameters. 
Hence, the singles equations retain their original form but the doubles amplitudes are approximated as
\begin{equation}
    \bra{\Phi_{ij}^{ab}}\tilde{H}+\left[\hat{F},\hat{T}_2\right]\ket{\Phi_0}=0\;.
\end{equation}\newline
To derive the working equations using the Cholesky representation, the two-electron integrals are substituted by the representation in equation~\ref{eq:Cholesky_Zweielektronenmatrix_MO_Basis} followed by reordering the contractions with the {intent} to
\begin{enumerate}
    \item eliminate the storage of intermediates which have three or more virtual indices (e.g. $VVVO$ and $VVVV$).%
    \item identify the optimal factorization by utilizing amplitude-transformed Cholesky vectors.%
\end{enumerate}%
In the following a generalization of the formulation in Refs.~\onlinecite{Krylov.2013} and \onlinecite{GaussJ.Chem.Phys.2019} to the ff regime is presented. 
Here, care has to be taken as the symmetry $L_{\mu\nu}^J=L_{\nu\mu}^{J*}$ is lifted due to the loss of eightfold permutational symmetry. Additionally, an antisymmetric formulation of the intermediates is chosen which is exploited during the contraction with the amplitudes in our implementation. Regarding the underlying ff-(EOM)-CCSD equations we refer to Refs.~\onlinecite{Stopkowicz.2015} and \onlinecite{Hampe.J.Chem.Phys.2017}. For the intermediates and amplitude-transformed Cholesky vectors the notation of Epifanovsky et al.\cite{Krylov.2013} was adopted: The numbers 1, 2, or 3 in the upper index mark a contraction with $L_{oo}^J$, $L_{ov}^J$ and $L_{vv}^J$ vectors with either a CC amplitude $T$, or a right/left EOM-CC amplitude $R$/$L$. $o$ and $v$ stand for the occupied and virtual space, respectively. The expressions are listed in Tabs.~\ref{tab:transformed_cholvec}, \ref{tab:ccsd_intermediates}, and \ref{tab:r_transformed_cholvec}. The connection between the choice of the intermediates and a diagrammatic approach is described in the appendix.

\begin{table}[tb]
{\centering
\caption{$t$-transformed Cholesky vectors}
\label{tab:transformed_cholvec}
\begin{ruledtabular}
\begin{tabular*}{0.48\textwidth}{@{\extracolsep{\fill}}lcllcl}
     $M_J^{2T}$&=&$\sum\limits_{me}L_{me}^{J}t_m^{e}$& $M_J^{2T*}$&=&$\sum\limits_{me}L_{em}^{J*}t_m^{e}$\\
     
     $M_{aiJ}^{1T}$&=&$\sum\limits_m{L_{mi}^{J}t_m^{a}}$& $M_{iaJ}^{1T*}$&=&$\sum\limits_m{L_{im}^{J*}t_m^{a}}$\\
      
     $M_{aiJ}^{2T}$&=&$\sum\limits_{me}L_{me}^{J}t_{im}^{ae}$& $M_{iaJ}^{2T*}$&=&$\sum\limits_{me}L_{em}^{J*}t_{im}^{ae}$\\ 
     
    $M_{aiJ}^{3T}$&=&$\sum\limits_fL_{af}^Jt_i^{f}$& $M_{iaJ}^{3T*}$&=&$\sum\limits_fL_{fa}^{J*}t_i^{f}$\\
       
    $M_{aiJ}^{2TT}$&=&$\sum\limits_{em}L_{me}^{J}t_i^{e}t_m^{a}$& $M_{iaJ}^{2TT*}$&=&$\sum\limits_{em}L_{em}^{J*}t_i^{e}t_m^{a}$\\
      
    $M_{ij}^{J}$&=&$\sum\limits_eL_{ie}^Jt_j^{e}+L_{ij}^J$& $M_{ji}^{J*}$&=&$\sum\limits_eL_{ei}^{J*}t_j^{e}+L_{ji}^{J*}$ \\  
       
    $M_{ab}^{J}$&=&$L_{ab}^J-\sum\limits_mL_{mb}^Jt_m^{a}$ & $M_{ba}^{J*}$&=&$L_{ba}^{J*}-\sum\limits_mL_{bm}^{J*}t_m^{a}$  \\ 
   
    $M_{ai}^{J,t_1}$\textcolor{blue}{$^a$} 
    &=& $M_{aiJ}^{3T}-M_{aiJ}^{1T}-M_{aiJ}^{2TT}+L_{ai}^J$\\[3pt]
    $M_{ia}^{J*,t_1}$&=&$M_{iaJ}^{3T*}-M_{iaJ}^{1T*}-M_{iaJ}^{2TT*}+L_{ia}^{J*}$\\[3pt]
    $M_{ai}^J$&=& $M_{ai}^{J,t_1}+M_{aiJ}^{2T}$ &  $M_{ia}^{J*}$&=&$M_{ia}^{J*,t_1}+M_{iaJ}^{2T*}$\\
    $M_{iaJ}^{2\tau*}$&=&$\sum\limits_{me}L_{me}^{J}\tau_{im}^{ae}$&  $\tau_{ij}^{ab}$&=&$t_{ij}^{ab}+\mathcal{P}_{ij}^{-}t_i^at_j^b$\\ 
\end{tabular*}%
\end{ruledtabular}}
{\textcolor{blue}{$^a$}\footnotesize This is equivalent to the notation for the $t_1$-transformed Hamiltonian $M_{ai}^{J,t_1}=\sum_{\mu\nu}\hat{C}_{\mu a}^*L_{\mu\nu}^J\hat{C}_{\nu i}$ with $\hat{C}_{\mu a}^*=C_{\mu a}^*-\sum_k C_{\mu m}^* t_m^a$ and $\hat{C}_{\mu i}=C_{\mu i}+\sum_e C_{\mu e} t_i^e$ introduced in Ref.~\onlinecite{Hattig.J.Chem.Phys.2000}.}
\end{table}
The working equations for the CCSD single and double amplitudes are given as:
 \begin{equation}
    \begin{split}
        t_i^{a}\Delta_i^{a}
        &=f_{ai}+\sum_{e}f_{ae}t_i^{e}(1-\delta_{ae})\\
        &-\sum_{Jm}M_{mi}^{J}\left(M_{maJ}^{2T*}+M_{maJ}^{3T*}-M_{maJ}^{2TT*}\right)+\sum_JM_J^{2T*}M_{ai}^{J}\\
        &-\sum_{m}\left(\sum_{e}f_{me}t_i^{e}+f_{mi}(1-\delta_{mi})\right)t_m^{a}+\sum_{Je}M_{ae}^{J}M_{ieJ}^{2T*}\\
        &+\sum_{me}t_{im}^{ae}\left[f_{me}-\sum_{Jn}\left(M_{mn}^{J}-L_{mn}^{J}\right)L_{en}^{J*}\right]\;,
    \end{split} 
\end{equation}
\begin{equation}
    \begin{split}
        t_{ij}^{ab}\Delta_{ij}^{ab}=&W_{abij}^{(t)}+\mathcal{P}_{ab}^{-}\Biggl\{\sum_{J}M_{ai}^JM_{jb}^{J*}+\sum_{f}t_{ij}^{af}(F_{bf}-f_{bf}\delta_{bf})\\
        &+\mathcal{P}_{ij}^{-}\left[\sum_{me}\left(\frac{1}{2}W_{mbei}^{(1t)}+\sum_JM_{mi}^{J}M_{eb}^{J*}\right)t_{jm}^{ae}\right]\Biggr\}\\
        &-\mathcal{P}_{ij}^{-}\sum_{m}t_{im}^{ab}(F_{mj}-f_{mj}\delta_{mj})+\frac{1}{2}\sum_{mn}W_{mnij}^{(t)}t_{mn}^{ab}\;.
    \end{split}
\end{equation}
{With the definitions $\Delta_i^{a}=f_{ii}-f_{aa}$ and $\Delta_{ij}^{ab}=f_{ii}+f_{jj}-f_{aa}-f_{bb}$, which equate to the orbital-energy difference for canonical orbitals, i.e., $\Delta\epsilon_i^{a}=\epsilon_{i}-\epsilon_{a}$ and  $\Delta\epsilon_{ij}^{ab}=\epsilon_{i}+\epsilon_{j}-\epsilon_{a}-\epsilon_{b}$.}
Within the CC2 scheme, in iteration $n$ the contributions $t_1^{(n-1)}\rightarrow t_1^{(n)}$ and $t_1^{(n-1)}\rightarrow t_2^{(n)}$ are calculated first and afterwards the immediate back contributions of the doubles $t_2^{(n)}\rightarrow t_1^{(n)}$ are accounted for similar to the discussion in Ref.~\onlinecite{Matthews2015}. While the singles equation retain their original form the doubles amplitudes simplify by keeping only the $t_1$ contributions:
\begin{equation}
     t_{ij}^{ab}\Delta{{\epsilon}_{ij}^{ab}} = \mathcal{P}_{ab}^{-}\sum_{J} M_{ai}^{J,t_1} M_{jb}^{J*,t_1}\;.
\end{equation}
The final energy expression is 
\begin{equation}
    E_{\text{CC}}=\sum_{ia}f_{ia}t_{i}^{a}+\frac{1}{2}\sum_{ia}\sum_{J}L_{ia}^JM_{iaJ}^{2\tau*}\;.
\end{equation}
Considering only the $\mathcal{O}(N^6)$ terms which have to be calculated in every iteration this formulation of the CCSD equations (while neglecting spin and symmetry) scales as
\begin{equation}
    V^4O^2+2V^3O^3+2V^2O^4\;
\end{equation}
for building the intermediates $W^{(t)}_{abij}$, $W^{(1t)}_{iabj}$, and $W^{(t)}_{ijkl}$ as well as the subsequent contraction of the latter two intermediates. Here, $O$ and $V$ correspond to the number of occupied and virtual orbitals, respectively. As discussed already in Ref.~\onlinecite{Krylov.2013}, in comparison to a canonical CCSD implementation, one less $O^3V^3$ type contraction is required. 
Hence, in theory the CD implementation has the same scaling with a slightly smaller prefactor than the canonical algorithm. 
Still, the expensive particle-particle ladder term does not factorize well. For its implementation special care has to be taken to not fully store the $VVVV$ intermediate emerging from rebuilding partially transformed integrals via $\sum_{J}M_{af}^{J}M_{eb}^{J*}$ which is covered in more detail in Sec.~\ref{sec:implementation}. We note, however, that in practical applications the $N^5$ term ($V^4N_{\text{CH}}$) often scales worse than the particle-particle ladder term ($V^4O^2$) itself, when $N_{\text{CH}}>O^2$. Nevertheless the implementation allows CCSD calculations at comparable timings but with significantly reduced memory requirements.\newline
CD-CC2 factorizes very well and scales only with $2O^2V^2N_{\text{CH}}$ FLOPs compared to canonical CC2 which scales as $OV^4+8O^2V^3+8O^3V^2+O^4V$. Additionally, no $VVVO$ or larger intermediate is needed at all, leading to much reduced memory demands.

\begin{table*}[tb]
\centering
\caption{List of (EOM)-CCSD intermediates}
\label{tab:ccsd_intermediates}
\begin{ruledtabular}
\begin{tabular*}{\textwidth}{@{\extracolsep{\fill}}lclc}
\multicolumn{3}{c}{Equation} & Scaling \\
\midrule
\midrule  
\multicolumn{4}{c}{CCSD}  \\
\midrule 
 $F_{ab}$&=&$f_{ab}-\sum\limits_{m}f_{mb}t_{m}^{a}-\sum\limits_{Jm}L_{mb}^{J}\left(M_{maJ}^{2T*}+M_{maJ}^{3T*}-M_{maJ}^{2TT*}\right)+\sum\limits_{J}M_{J}^{2T}M_{ba}^{J*}$&\\
 $F_{ij}$&=&$f_{ij}+\sum\limits_{e}f_{ie}t_{j}^{e}+\sum\limits_{Je}M_{ejJ}^{2T}L_{ei}^{J*}+\sum\limits_{J}M_{J}^{2T}M_{ji}^{J*}-\sum\limits_{Jm}\left(M_{im}^{J}-L_{im}^{J}\right)M_{jm}^{J*}$&\\
 $F_{ia}$&=&$f_{ia}+\sum\limits_{J}L_{ia}^{J}M_{J}^{2T*}-\sum\limits_{Jm}\left(M_{im}^{J}-L_{im}^{J}\right)L_{am}^{J*}$&\\
 $W_{iabj}^{(1t)}$&=&$\sum\limits_{me}t_{jm}^{ae}\sum\limits_{J}L_{ie}^{J}L_{bm}^{J*}$&$V^3O^3$\\
 $W_{iabj}^{(2t)}$&=&$\sum\limits_{me}t_{jm}^{ae}\sum\limits_{J}L_{mb}^{J}L_{ei}^{J*}+\sum\limits_JM_{ij}^{J}M_{ba}^{J*}$& $V^3O^3$\\
 $W_{abij}^{(t)}$&=&$-\frac{1}{2}\sum\limits_{ef}\left(\mathcal{P}_{ab}^{-}\sum\limits_{J}M_{af}^{J}M_{eb}^{J*}\right)t_{ij}^{ef}$& $V^4O^2$\\
 
 $W_{ijkl}^{(t)}$&=&$\;\;\;\frac{1}{2}\sum\limits_{ef}\left(\mathcal{P}_{ef}^{-}\sum\limits_{J}L_{ie}^{J}L_{fj}^{J*}\right)t_{kl}^{ef}+\mathcal{P}_{kl}^{-}\sum\limits_JM_{ik}^JM_{lj}^{J*}$&$V^2O^4$\\
 $W_{kaij}^{(t)}$&=&$-\mathcal{P}_{ij}^{-}\Biggl(\sum\limits_{J}M_{kj}^JM_{ia}^{J*}+\sum\limits_{en}t_{jn}^{ae}\sum\limits_{J}L_{ke}^{J}M_{in}^{J*}\Biggr)+\frac{1}{2}\sum\limits_{ef}t_{ij}^{ef}\left(\mathcal{P}_{ef}^{-}\sum\limits_{J}L_{ke}^{J}M_{fa}^{J*}\right)-\sum\limits_{f}t_{ij}^{af}F_{kf}$&$V^3O^3+V^2O^4$\\
\midrule  
\multicolumn{4}{c}{EOM-EE}  \\
\midrule 
 $T_{ab}$&=&$\sum\limits_{Jm}L_{mb}^{J}\left(M_{maJ}^{2R*}+M_{maJ}^{3R*}-M_{maJ}^{2RT*}\right)-\sum\limits_{J}M_{ab}^{J}M_{J}^{2R*}$&\\  
 $T_{ij}$&=&$\sum\limits_{J}L_{ie}^{J}M_{jeJ}^{2R*}+\sum\limits_{J}M_{ij}^{J}M_{J}^{2R*}-\sum\limits_{Jm}M_{mj}^{J}M_{miJ}^{2R*}$&\\
 $W_{ijkl}^{(r)}$&=&$\;\;\;\frac{1}{2}\sum\limits_{ef}\left(\mathcal{P}_{ef}^{-}\sum\limits_{J}L_{ie}^{J}L_{fj}^{J*}\right)r_{kl}^{ef}$&$V^2O^4$\\
 $W_{abij}^{(r)}$&=&$-\frac{1}{2}\sum\limits_{ef}\left(\mathcal{P}_{ab}^{-}\sum\limits_{J}M_{af}^{J}M_{eb}^{J*}\right)r_{ij}^{ef}$& $V^4O^2$\\
 $W_{abij}^{(l)}$&=&$-\frac{1}{2}\sum\limits_{ef}\left(\mathcal{P}_{ab}^{-}\sum\limits_{J}M_{fa}^{J}M_{be}^{J*}\right)l_{ef}^{ij}$&$V^4O^2$\\
\end{tabular*}%
\end{ruledtabular}
\end{table*}

%%%%%%%%%%%%%%%%%%%%%%%%%%%%%%%%%%%%%%
\subsection{Equation-of-Motion-Coupled-Cluster Theory}\label{sec:EOMCoupledCluster}
%%%%%%%%%%%%%%%%%%%%%%%%%%%%%%%%%%%%%%

In the EOM-CC approach\cite{Shavitt2009.MBPT, Bartlett.Chem.Phys.Lett.1989, Stanton.J.Chem.Phys.1993, Levchenko.J.Chem.Phys.2004, Krylov.Acc.Chem.Res.2005, Stanton.J.Chem.Phys.1994, Bartlett.J.Chem.Phys.1995, Sneskov2011, Hampe.J.Chem.Phys.2017, Hampe.2019.Dissertation, Petros.2023.Dissertation}, the target state is obtained from a CC reference wave function 
\begin{equation}
    \ket{\Psi_{\text{exc}}}=\sum_n\hat{R}_n\ket{\Psi_{\text{CC}}}=\hat{R}\text{e}^{\hat{T}}\ket{\Phi_0}
\end{equation}
with the linear operator
\begin{equation}
    \hat{R}=\hat{R}_0+\hat{R}_1+\hat{R}_2+\hdots+\hat{R}_N=\sum_I r_I \hat{\mu}_I\;.
\end{equation}
$\hat{\mu}_I$ generates the excited determinants $\Phi_I$ of interest and $\hat{R}$ is defined for the particle conserving electronically-excited and spin-flip states (EOM-EE\cite{Bartlett.Chem.Phys.Lett.1989, Stanton.J.Chem.Phys.1993}, EOM-SF\cite{Levchenko.J.Chem.Phys.2004, Krylov.Acc.Chem.Res.2005}), as
\begin{equation}
    \hat{R}_n^{\text{EE/SF}}=\left(\frac{1}{n!}\right)^2\sum_{i,j\ldots}\sum_{a,b\ldots}r_{ij\ldots}^{ab\ldots}\hat{a}_a^\dagger\hat{a}_i\hat{a}_b^\dagger\hat{a}_j\ldots
\end{equation}
as well as for ionized (EOM-IP\cite{Stanton.J.Chem.Phys.1994})
\begin{equation}
    \hat{R}_n^{\text{IP}}=\left(\frac{1}{n!(n-1)!}\right)\sum_{i,j\ldots k}\sum_{a,b\ldots}r_{ij\ldots k}^{ab\ldots}\hat{a}_a^\dagger\hat{a}_i\hat{a}_b^\dagger\hat{a}_j\ldots \hat{a}_k
\end{equation}
and electron-attached states (EOM-EA\cite{Bartlett.J.Chem.Phys.1995})
\begin{equation}
    \hat{R}_n^{\text{EA}}=\left(\frac{1}{n!(n-1)!}\right)\sum_{i,j\ldots}\sum_{a,b\ldots c}r_{ij\ldots}^{ab\ldots c}\hat{a}_a^\dagger\hat{a}_i\hat{a}_b^\dagger\hat{a}_j\ldots\hat{a}_c^\dagger\;.
\end{equation}
The EOM-CC eigenvalue equations are given as
\begin{equation}\label{eq:EOM-CC-Gleichungen}
    \begin{split}
      \bra{\Phi_I}\left[\bar{H},\hat{R}\right]\ket{\Phi_0}&=E_{\text{exc}}\bra{\Phi_I}\hat{R}\ket{\Phi_0}\\
      \sum_J\bra{\Phi_I}(\bar{H}\hat{\mu}_J)_c\ket{\Phi_0}r_J&=E_{\text{exc}}r_I\\
      \mathbf{A}\mathbf{r}&=E_{\text{exc}}\mathbf{r}
   \end{split}
\end{equation}
where the eigenvalue $E_{\text{exc}}$ is the excitation energy, $\mathbf{A}$ is the CC Jacobian\cite{Sneskov2011}, and the eigenvectors $\mathbf{r}$ contain the amplitudes of the operator $\hat{R}$. 
The matrix-vector product of the Jacobian with the EOM-CC amplitude vector is the sigma vector, $\mathbf{\sigma}=\mathbf{A}\mathbf{r}$, thereby referring to the usual notation of the Davidson algorithm.\cite{Davidson1975,Hirao1982,Caricato2010} Truncation of $\hat{T}$ and $\hat{R}$ at the same excitation level (here doubles) yields:
\begin{alignat}{4}
    E_{\text{exc.}}r_0&=\bra{\Phi_0}(\Bar{H}\hat{R}_1)_c\ket{\Phi_0} &+&\bra{\Phi_0}(\Bar{H}\hat{R}_2)_c\ket{\Phi_0}\;,\\
    E_{\text{exc.}}r_{i}^{a}&=\bra{\Phi_{i}^{a}}(\Bar{H}\hat{R}_1)_c\ket{\Phi_0} &+&\bra{\Phi_{i}^{a}}(\Bar{H}\hat{R}_2)_c\ket{\Phi_0}\;,\\
    E_{\text{exc.}}r_{ij}^{ab}&=\bra{\Phi_{ij}^{ab}}(\Bar{H}\hat{R}_1)_c\ket{\Phi_0} &+&\bra{\Phi_{ij}^{ab}}(\Bar{H}\hat{R}_2)_c\ket{\Phi_0}\;.
\end{alignat}
\begin{table}[tb]
\centering
\caption{$r$ and $l$-transformed Cholesky vectors}
\label{tab:r_transformed_cholvec}
\begin{ruledtabular}
\begin{tabular*}{0.48\textwidth}{@{\extracolsep{\fill}}lcllcl}
  \multicolumn{6}{c}{EOM-EE}  \\
 \midrule
 $M_J^{2R*}$&=&$\sum\limits_{me}L_{em}^{J*}r_m^{e}$ & $M_J^{2L*}$&=&$\sum\limits_{me}{ M}_{em}^{J*}l^m_{e}$\\
 $M_{iaJ}^{1R*}$&=&$\sum\limits_m{L_{im}^{J*}r_m^{a}}$ & $M_{aiJ}^{1L*}$&=&$\sum\limits_m{{ M}_{mi}^{J*}l^m_{a}}$\\
 $M_{iaJ}^{2R*}$&=&$\sum\limits_{me}L_{em}^{J*}r_{im}^{ae}$ & $M_{aiJ}^{2L*}$&=&$\sum\limits_{me}{ M}_{me}^{J*}l^{im}_{ae}$\\
 $M_{iaJ}^{3R*}$&=&$\sum\limits_eL_{ea}^{J*}r_i^{e}$ & $M_{aiJ}^{3L*}$&=&$\sum\limits_e{ M}_{ae}^{J*}l^i_{e}$\\
 $M_{iaJ}^{2RT*}$&=&$\sum\limits_{me}L_{em}^{J*}r_i^{e}t_m^{a}$ & &&\\
% ${\color{red}\cancel{ M_{iaJ}^{2TR*} }}$&=&${\color{red}\cancel{\sum\limits_{me}L_{em}^{J*}t_i^{e}r_m^{a} }} $& &&\\
 $M_{jiJ}^{2R*}$&=& $\sum\limits_eL_{ei}^{J*}r_j^{e}$ & && \\   
\end{tabular*}%
\end{ruledtabular}
\end{table}
The working equations for EOM-EE ($\hat{R}=\hat{R}^{\text{EE}}$) in the CD formalism are given for the $r_0$ amplitude equations as
\begin{equation}
    E_{\text{exc.}}r_0=\sum\limits_{ia}F_{ia}r_{i}^{a}+\frac{1}{2}\sum_{ia}\sum_JL_{ia}^JM_{iaJ}^{2R*},
\end{equation}
as well as the for $r_1$ amplitudes as
\begin{equation}
    \begin{split}
        E_{\text{exc.}}r_{i}^{a}&=\sum_{e}F_{ae}r_i^{e}-\sum_{m}F_{mi}r_m^{a}+\sum_{me}F_{me}r_{im}^{ae}+\sum_{J} M_{ai}^{J}M_J^{2R*}\\
        &-\sum_{mJ}M_{mi}^{J}\left(M_{maJ}^{3R*}-M_{maJ}^{2RT*}+M_{maJ}^{2R*}\right)\\
        &-\sum_{em}t_{im}^{ae}\sum_{Jn}L_{ne}^{J}M_{nmJ}^{2R*}+\sum_{Je}M_{ae}^JM_{ieJ}^{2R*},\\
    \end{split}
\end{equation}
and for the $r_2$ amplitudes as
\begin{equation}
    \begin{split}
        E_{\text{exc.}} r_{ij}^{ab} & = W_{abij}^{(r)}+\frac{1}{2}\sum_{mn}W_{mnij}^{(t)}r_{mn}^{ab}+\mathcal{P}_{ij}^{-}\mathcal{P}_{ab}^{-}\Biggl\{\sum_{me}t_{jm}^{ae}\left(\sum_{J}M_{be}^{J}M_{imJ}^{2R*}\right)\\
        &+\sum_{J}M_{ai}^{J}\Biggl(M^{2R*}_{jbJ} - M^{2RT*}_{jbJ} + M^{3R*}_{jbJ}\Biggr)+\sum_{me}r_{jm}^{ae}W_{mbei}^{(2t)}\Biggr\}\\
        &+ \mathcal{P}_{ab}^{-} \Biggl( \sum_{e}T_{ae}t_{ij}^{be}+\sum_{e}F_{be}r_{ij}^{ae}-\sum_{m}W_{mbij}^{(t)}r_{m}^{a}\Biggr)\\
        &+\mathcal{P}_{ij}^{-}\Biggl(\sum_{m}r_{jm}^{ab}F_{mi}+\sum_{m}t_{jm}^{ab}\left(\sum_{e}F_{me}r_{i}^{e}+T_{mi}\right)\Biggr)\\
        &+\frac{1}{2}\sum_{mn}t_{mn}^{ab}\left(\mathcal{P}_{ij}^{-}\mathcal{P}_{mn}^{-}\sum_{J}M_{mi}^{J}M_{jnJ}^{2R*}+W_{mnij}^{(r)}\right).
    \end{split}
\end{equation}
We note that compared to Ref.~\onlinecite{Krylov.2013} we redefined the following contribution:
\begin{equation}
    \begin{split}
        \text{\underline{Ref.~\onlinecite{Krylov.2013}}}\qquad &\hphantom{=} \qquad\text{\underline{This work}} \\
       \sum_{nf}t^{bf}_{in}W_{nafj}^{(1r)}&=\sum_{me}r_{jm}^{ae}W_{mbei}^{(1t)}\\
    \sum_{nf}t^{bf}_{in}\left(\sum_{me}\sum_{J}L_{mf}^{J}L_{en}^{J*}r_{jm}^{ae}\right)&=
   \sum_{me}r_{jm}^{ae}\left(\sum_{nf}\sum_{J}L_{mf}^{J}L_{en}^{J*}t^{bf}_{in}\right)
    \end{split}
\end{equation}
so that the intermediate $W_{mbei}^{(1t)}$ is independent of the EOM-CC amplitudes in contrast to $W_{nafj}^{(1r)}$. Grouping of the former with $I_{abij}^{(3i)}$ yields 
\begin{equation}
        \mathcal{P}_{ab}^{-}\mathcal{P}_{ij}^{-}\sum_{me}r_{jm}^{ae}W_{mbei}^{(2t)}=\mathcal{P}_{ab}^{-}\mathcal{P}_{ij}^{-}\sum_{me}r_{jm}^{ae}\left(W_{mbei}^{(1t)}+\sum_JM_{mi}^{J}M_{eb}^{J*}\right).
\end{equation}
This proves to be more efficient since another $V^3O^3$ type contraction does not need to be re-evaluated in each iteration. In total, the scaling corresponds to $V^4O^2+2V^3O^3+3V^2O^4$. In comparison, the canonical implementation scales as $V^4O^2+V^3O^3+V^2O^4$. Thus, the Cholesky implementation comes with some drawback in terms of efficiency as compared to the canonical case due to the fact that $r$-dependent intermediates such as $W_{ijkl}^{(r)}$ (responsible for $2V^2O^4$) and $\sum_{J}M_{be}^{J}M_{imJ}^{2R*}$ (responsible for $V^3O^3$) and their contractions with the cluster amplitudes have to be recomputed in each iteration to avoid the storage of $VVVV$ and $VVVO$ type intermediates, respectively. 
Likewise, the comments made for the $W_{abij}^{(t)}$ intermediate for ground state CC hold here as well for $W_{abij}^{(r)}$.

%%%%%%%%%%%%%%%%%%%%%%%%%%%%%%%%%%%%%%%%%%%%%%%%%%%%%%%%%%%%%%
\subsubsection{Approximate EOM-Coupled-Cluster - EOM-EE-CC2}
%%%%%%%%%%%%%%%%%%%%%%%%%%%%%%%%%%%%%%%%%%%%%%%%%%%%%%%%%%%%%%

As seen for the EOM-CCSD equations a straightforward way to derive the EOM-CC2\cite{Christiansen.Chem.Phys.Lett.1995, Christiansen.J.Chem.Phys.1995, Koch.J.Chem.Phys.1997, Hattig.J.Chem.Phys.2000, Kitsaras.J.Chem.Phys.2024} equations is to evaluate the Jacobian, i.e., the derivative of the CC2 amplitude equations with respect to the cluster amplitudes. For canonical orbitals the CC2-Jacobian is given as
\begin{equation}\label{eq:CC2-Jacobian}
\begin{split}
        &A_{IJ}=\pdv[]{\Omega_I(\boldsymbol{t})}{t_J}=\begin{pmatrix}
        \pdv[]{\Omega_1(\boldsymbol{t})}{t_1} & \pdv[]{\Omega_1(\boldsymbol{t})}{t_2} \\
        \pdv[]{\Omega_2(\boldsymbol{t})}{t_1} & \pdv[]{\Omega_2(\boldsymbol{t})}{t_2}
    \end{pmatrix}\\
    =&\begin{pmatrix}
        \bra{\Phi_1}(\tilde{H}\hat{\mu}_1)_c+((\tilde{H}\hat{\mu}_1)_c\hat{T}_2)_c\ket{\Phi_0} & \bra{\Phi_1}(\tilde{H}\hat{\mu}_2)_c\ket{\Phi_0} \\
        \bra{\Phi_2}(\tilde{V}\hat{\mu}_1)_c\ket{\Phi_0} & {-}\Delta\epsilon_2
    \end{pmatrix}.
\end{split}
\end{equation}
While the determining equations for the singles amplitudes are the same as in the full CCSD treatment, the doubles amplitudes are simply given as
 \begin{equation}
     E_{\text{exc}}r_{ij}^{ab}=\bra{\Phi_{ij}^{ab}}(\tilde{V}\hat{R}_1)_c\ket{\Phi_0}{-}\Delta\epsilon_{ij}^{ab}r_{ij}^{ab}
\end{equation}
or rather
\begin{equation}
    r_{ij}^{ab}=\frac{\bra{\Phi_{ij}^{ab}}(\tilde{V}\hat{R}_1)_c\ket{\Phi_0}}{E_{\text{exc}}{+}\Delta\epsilon_{ij}^{ab}}\;.
\end{equation}
Within the CD approximation the EOM-EE-CC2 working equations are
\begin{equation}
    \begin{split}
          (E_{\text{exc.}}{+}\Delta\epsilon_{ij}^{ab})r_{ij}^{ab}   &=  \mathcal{P}_{ij}^{-}\sum_{m}r_{jm}^{ab}f_{mi} +  \mathcal{P}_{ab}^{-}\sum_{e}f_{be}r_{ij}^{ae}\\
                         & +\mathcal{P}_{ab}^{-}\mathcal{P}_{ij}^{-}\sum_{J} M_{ai}^{J,t1} \Biggl(M_{jbJ}^{3R*}   \\ 
                         & {- M_{jbJ}^{2RT*} -\sum_m M_{jm}^{J*}r_{m}^{b} }\Biggr).
    \end{split}
\end{equation}
In analogy to CD-EOM-CCSD the scaling of CD-EOM-CC2 is $2O^2V^2N_{\text{CH}}+O^2V^3+O^3V^2$ which is more expensive than the canonical implementation, i.e., $3O^2V^3+3O^3V^2$, for $O+V<N_{\text{CH}}$. As an advantage, the CD scheme does not require building any four-index intermediates. Thus the memory is in principle only limited by the storage of the amplitudes themselves. {To be more precise, the EOM-CC2 amplitudes for double excitations can be partitioned out of the Davidson subspace. This allows to form the amplitudes temporarily and process them in batches. Consequently for CC2, the primary limitation on memory usage arises from the Cholesky vectors.}
%

%%%%%%%%%%%%%%%%%%%%%%%%%%%%%%%%%%%%%%%%%%%%%%%%%%%%%%%%%%
\subsection{Left-hand side EOM-CC Equations and Properties}\label{sec:LambdaEquations}
%%%%%%%%%%%%%%%%%%%%%%%%%%%%%%%%%%%%%%%%%%%%%%%%%%%%%%%%%%

Since the Jacobian $\bm{A}$ in equation~\ref{eq:EOM-CC-Gleichungen} is a non-Hermitian matrix (due to the fact that the exponential of the cluster operator $e^{\hat{T}}$ is non-unitary), there exists also a left-hand side equation of the eigenvalue problem\cite{Shavitt2009.MBPT,Sneskov2011}
\begin{equation}\label{eq:EOM_Eigenvalue_problem_lefthandside}
\bm{l}^T \bm{A}=E_{\text{exc}} \bm{l}^T
\end{equation}
to the same set of eigenvalues. These equations similarly describe linear deexciations via the operator 
\begin{equation}
    \hat{L}=\hat{L}_0+\hat{L}_1+\hat{L}_2+\hdots+\hat{L}_N=\sum_I l_I \hat{\mu}_I^{\dagger}\;.
\end{equation}
In this case, the eigenvectors $\boldsymbol{r}$ and $\boldsymbol{l}$ are not simply Hermitian conjugates but form an biorthogonal set which can be normalized:
\begin{equation}
    \begin{aligned}
        \bra{ 0}\hat{L}^m \hat{R}^n\ket{0} =\sum_I l_I^m r_I^n & =\delta_{m n}\;.
    \end{aligned}
\end{equation}
As mentioned above, the left-hand side equation does not have to be solved to evaluate the excitation energies. It has to be solved for the evaluation of properties.\cite{Stanton.J.Chem.Phys.1993, Stanton1995, Shavitt2009.MBPT, GaussJ.Chem.Phys.2019, Hampe.J.Chem.TheoryComput.2019} For example, (transition) dipole moments, can be expressed as biorthogonal expectation value 
\begin{equation}
    \mu_{mn}=\bra{\Psi_{\mathrm{EOM}}^{m}}\hat{\mu} \ket{\Psi_{\mathrm{EOM}}^{n}}=\bra{\Phi_0}\hat{L}^{m} \mathrm{e}^{-\hat{T}} \hat{\mu} \mathrm{e}^{\hat{T}} \hat{R}^{n}\ket{\Phi_0}\;,
\end{equation}
neglecting the contributions from the cluster amplitudes and orbital relaxation.\cite{Handy1984}
For $m\neq n$, this yields transition dipole moments between the states $m$ and $n$ and single state dipole moments for $m=n$. A special case is $m=n=0$ which is equivalent to the CC reference wave function. The CC ground state is a solution to the eigenvalue problem of EOM-EE, where $r_0$ is set to 1 and all other amplitudes of higher excitations vanish. 
\begin{equation}
    \mu_{00}=\left\langle \Phi_0\left|\hat{L}^0 \bar{\mu}\right| \Phi_0\right\rangle
\end{equation}
Thus, the problem reduces to the EOM left-hand-side solution for the ground state. This is formally equivalent to the $\Lambda$-equations in CC derivative theory\cite{Scheiner1987,Gauss1991,Gauss.Molecular.Properties.2000,Shavitt2009.MBPT} with $\hat{L}^0=(1+\hat{\Lambda})$, i.e. $l_0=1$ for the ground state. 
 For excited states, $l_0$ can be chosen as 0, as the biorthogonality condition is ensured by scaling the rest of the $l$ vector.
The transition dipole moment between two arbitrary states is evaluated by summing over the dipole moment integrals $\mu$ and the reduced one-particle transition-density matrix $\rho$
\begin{equation}
    \mu_{mn}=\sum_{pq} \mu_{pq}\bra{\Phi_0}\hat{L}^{m} e^{-\hat{T}} \hat{a}_p^{\dagger}\hat{a}_q e^{\hat{T}} \hat{R}^{n}\ket{\Phi_0}=\sum_{pq} \mu_{pq}\rho_{pq}^{mn}\;.
\end{equation}
The evaluation of one electron reduced density matrix is unchanged in a Cholesky decomposed implementation. For reference of the working equations in the ff setting see Ref.~\onlinecite{Hampe.J.Chem.TheoryComput.2019}.
\begin{table}[tb]
\centering
\caption{List of left-hand side EOM-CCSD intermediates}
\label{tab:lambda_ccsd_intermediates}
\begin{ruledtabular}
    \begin{tabular*}{0.48\textwidth}{@{\extracolsep{\fill}}lcllcl}
     $\tilde{l}_{ij}$&=&$\;\;\;\frac{1}{2}\sum\limits_{efm}l_{ef}^{mi}t_{mj}^{ef}$ &
     $\tilde{l}_{ab}$&=&$-\frac{1}{2}\sum\limits_{mne}t_{mn}^{ae}l_{eb}^{mn}$\\
     $\tilde{l}_{i}$&=&$-\frac{1}{2}\sum\limits_{efm}l_{ef}^{m}t_{mi}^{ef}$   &    
     $\tilde{l}_{a}$&=&$\;\;\;\frac{1}{2}\sum\limits_{mne}t_{mn}^{ae}l_{e}^{mn}$\\
    \end{tabular*}%
\end{ruledtabular}
\end{table}
Consequentially the ground and excited-state properties require the solution of the EOM left-hand side equations. For CD-(EOM)-CCSD the $l_1$ amplitudes are given by
\begin{equation}
    \begin{split}
         E_{\mathrm{exc}}l^{i}_{a}     =  &F_{ia}l_0 + \sum_e F_{ea} l_{e}^{i} - \sum_m F_{im}l_{a}^{m} - \sum_{m} F_{ma} \tilde{l}_{im} \\
                       & - \sum_{me} W_{ieam}^{(2t)} l_{e}^{m} + \frac{1}{2}\sum_{mne} W_{ienm}^{(t)} l_{ae}^{mn} + \sum_{Je} M_{ea}^J M_{eiJ}^{2L*} \\
                       & + \sum_{J} L_{ia}^{J}\left(\sum_{me}M_{me}^{J*}l_{e}^{m} - \sum_{mn} M_{mn}^{J*}\tilde{l}_{mn} + \sum_{ef}M_{ef}^{J*}\tilde{l}_{ef}\right) \\
                       & + \sum_{Jn} L_{na}^{J} \Biggl( \sum_{m} M_{mi}^{J*} \tilde{l}_{mn} + \frac{1}{2} \sum_{mo} M_{mo}^{J*} \sum_{ef} l_{ef}^{im} t_{no}^{ef}  \\
                       & - \sum_{ge} M_{ge}^{J*} \sum_{mf} t_{nm}^{gf} l_{ef}^{im}\Biggr) - \sum_{Je} L_{ie}^{J} \sum_{f} M_{af}^{J*} \tilde{l}_{ef}
    \end{split}
\end{equation}
and $l_2$ amplitudes by
\begin{equation}
    \begin{split}
         E_{\mathrm{exc}}l^{ij}_{ab}   =  & W_{abij}^{(l)} - \mathcal{P}_{ab}^{-}\mathcal{P}_{ij}^{-}\sum_{me} W_{jebm}^{(2t)} l_{ae}^{im}\\
                         & +\mathcal{P}_{ab}^{-}\mathcal{P}_{ij}^{-} F_{ia}l_{b}^{j} + \mathcal{P}_{ab}^{-} \sum_{e} F_{eb} l_{ae}^{ij} +                            \mathcal{P}_{ij}^{-} \sum_{m} F_{im} l_{ab}^{jm} \\
                         & + \frac{1}{2} \sum_{mn}W_{ijmn}^{(t)}l_{ab}^{mn}+\frac{1}{4}\sum_{mn}\left(\mathcal{P}_{mn}^{-}\sum_{J}L_{ma}^{J}L_{bn}^{J*}\right)\sum_{ef}t_{mn}^{ef}l_{ef}^{ij} \\
                         & + \mathcal{P}_{ab}^{-}\mathcal{P}_{ij}^{-} \sum_J L_{ia}^{J} \Biggl( M_{bjJ}^{3L*} - M_{bjJ}^{1L*} + M_{bjJ}^{2L*} \\
                         & - \sum_{m} L_{bm}^{J*} \tilde{l}_{jm} - \sum_{e} L_{ej}^{J*} \tilde{l}_{eb} +\frac{1}{2} L_{bj}^{J*}l_0\Biggr)\;.
    \end{split}
\end{equation}

%%%%%%%%%%%%%%%%%%%%%%%%%%%%%%%%%%%%%%%%%%%%%%%%%%%%%%%%%%%%
\subsubsection{Approximate EOM-Coupled-Cluster - EOM-EE-CC2}
%%%%%%%%%%%%%%%%%%%%%%%%%%%%%%%%%%%%%%%%%%%%%%%%%%%%%%%%%%%%
%
For CC2 the left-hand side equations are evaluated using the same strategy as for the right-hand side. According to the left-hand side eigenvalue problem in Eq.~\ref{eq:EOM_Eigenvalue_problem_lefthandside}, multiplication with the CC2 Jacobian of Eq.~\ref{eq:CC2-Jacobian} yields the equations
\begin{equation}
    E_{\mathrm{exc}}l_{a}^{i}=\bra{\Phi_0}(\hat{L}_2\tilde{V})_c\ket{\Phi_{i}^{a}}+\bra{\Phi_0}(\hat{L}_1\tilde{H})_c+((\hat{L}_1\tilde{H})_c)\hat{T}_2)c\ket{\Phi_{i}^{a}}
\end{equation}
and 
\begin{equation}
    l_{ab}^{ij}=\frac{\bra{\Phi_0}(\hat{L}_1\tilde{H})_c\ket{\Phi_{ij}^{ab}}}{ E_{\mathrm{exc}}{+}\Delta\epsilon_{ij}^{ab}}\;.
\end{equation}
Thus, for the left-hand side also the singles equations are altered by the CC2 approximation. Within the CD approximation the working equation are
\begin{equation}
    \begin{split}
        E_{\mathrm{exc}}l^{i}_{a}   &=  F_{ia}l_0 + \sum_e F_{ea} l_{e}^{i} - \sum_m F_{im}l_{a}^{m}\\
        &- \sum_{me} W_{ieam}^{(2t)} l_{e}^{m} + \sum_{J} M_{J}^{2L} L_{ai}^{J*}\\
        &+ \sum_{Jf} M_{fa}^{J} \sum_{me} M_{me}^{J*,t_1} l_{fe}^{im} - \sum_{Jn}  M_{in}^{J} \sum_{me} M_{me}^{J*,t_1} l_{ae}^{nm}
    \end{split}
\end{equation}
and    
\begin{equation}
    \begin{split}
         \left(E_{\mathrm{exc}}{+}\Delta\epsilon_{ij}^{ab}\right)l^{ij}_{ab}  &= \mathcal{P}_{ab}^{-}\mathcal{P}_{ij}^{-} F_{ia}l_{b}^{j} \\
         &+ \mathcal{P}_{ab}^{-} \sum_{e} f_{eb} l_{ae}^{ij} + \mathcal{P}_{ij}^{-} \sum_{m} f_{im} l_{ab}^{jm}\\
         &+ \mathcal{P}_{ab}^{-}\mathcal{P}_{ij}^{-} \sum_J L_{ia}^{J} \left( M_{bjJ}^{3L*} - M_{bjJ}^{1L*} +\frac{1}{2} L_{bj}^{J*}l_0\right)\;.
    \end{split}
\end{equation}
Thereby the scaling is identical to the right-hand side equations. Accordingly, the arguments given in the previous section hold here as well.
%

%%%%%%%%%%%%%%%%%%%%%%%%%%%%%%%%%%%%%%
\subsection{Implementation}\label{sec:implementation}
%%%%%%%%%%%%%%%%%%%%%%%%%%%%%%%%%%%%%%
%
Spin-integrated open-shell implementations of ground-state CCSD and CC2 as well as excited-state methods including EOM-CCSD (EE, SF, IP, EA) and the approximate EOM-CC2 (EE) method were implemented within the program package QCUMBRE\cite{QCUMBRE}. QCUMBRE is based on a string-based tensor contraction framework presented in Ref.~\onlinecite{Hampe.J.Chem.Phys.2017}. A contraction is generally performed by casting it into an efficient matrix multiplication by tensor transpositions which is evaluated using BLAS\cite{BLAS}. As the initial framework solely considered spin-symmetry, further development by Kitsaras\cite{Petros.2023.Dissertation} enabled the exploitation of point-group symmetry via symmetry-blocked tensors.\newline
In the following section, further improvements that have been added as part of this work are introduced. These improvements have been designed to enable the use of contractions in a more 'black box' manner while aiming at maintaining a developer-friendly structure that allows straightforward implementation of working equations:\newline
a) Transpositions, despite scaling only with the tensor size, noticeably affect the total wall time due to their inherent cache inefficiency. Both input tensors require transposition to ensure the correct shape for the matrix multiplication step, and the output has to be transposed as well to obtain the target quantity. Consequently, to perform a tensor contraction, in the worst case, three tensor transpositions have to be performed. 
As the indices for a given contraction may align with the target indices for any of the transpositions the number of FLOPs required is dependent on whether the tensor is the left-hand side or right-hand side input of the contraction. 
To achieve an optimal and black-box contraction routine, it was extended by a mechanism that determines the order of the input tensors by minimizing a cost function.\newline
b) To further increase the efficiency, antisymmetry of the tensors is exploited fully, similarly to the ideas discussed in Refs.~\onlinecite{Solomonik2014, Epifanovsky.J.Comput.Chem.2013}. The coupled-cluster amplitudes and intermediates are antisymmetric with respect to permutation, i.e.,
\begin{equation}
    t^{ab}_{ij}=-t^{ba}_{ij}=t^{ba}_{ji}=-t^{ab}_{ji}\;.
\end{equation}
Thus, for any such quantity it is sufficient to save the block $a<b$ and $i<j$. Exploiting this symmetry during the contractions such as in the particle-particle ladder term
\begin{equation}
    t_{i<j}^{a<b}\leftarrow2\sum_{e<f} W^{(t)}_{abef}t_{ij}^{ef}
\end{equation}
the required FLOPs are reduced by a factor of 6. Nonetheless, some care has to be taken for intermediates in which the tensors have to be partially (Eq.~\ref{eq:example_partially_unpack}) or fully (Eq.~\ref{eq:example_fully_unpack}) unpacked to perform the contraction:
\begin{equation}\label{eq:example_partially_unpack}
 t_{ij}^{a<b}\leftarrow -\mathcal{P}^-_{ij}\sum_{m}t_{im}^{ab}F_{mj}\;,
\end{equation}
\begin{equation}\label{eq:example_fully_unpack}
 t_{ij}^{ab}\leftarrow \frac{1}{2}\mathcal{P}^-_{ab}\mathcal{P}_{ij}^{-}\sum_{em}W_{mbei}^{(1t)}t_{jm}^{ae}\;.
\end{equation}
c) Additionally, as is standard practice for tensor contraction libraries, a parallelization of the most time-consuming steps in the tensor contraction workflow has been carried out. Thus, for QCUMBRE the tensor transpositions as well as the folding and unfolding of packed indices which precede the matrix multiplication step were parallelized. Furthermore, the \textsc{zgemm} routines employed for matrix multiplication are easily parallelized via OpenMP\cite{OpenMP} using the respective threaded BLAS libraries. 
In addition to the shared memory parallelization of the tensor contractions a massively parallel implementation of EOM using the message passing interface (MPI) was realized. As each root of EOM is independent of the others it is possible to calculate them on different nodes.
For a review of the developments for established quantum chemical packages handling field-free calculations we refer to the review~\onlinecite{Calvin2020}.\newline
d) For a memory-efficient implementation of CD-(EOM)-CCSD special care has to be taken to avoid the storage of any rank four tensor with three or more virtual indices e.g. $VVVO$ and $VVVV$. The particle-particle-ladder-like terms $W_{abij}^{(t/r/l)}$ as well as the EOM intermediate $W^{(t)}_{kaij}$ need such quantities if implemented in a straightforward manner. In order to avoid the storage, an on-the-fly three tensor contraction routine was implemented that builds these contributions batch-wise, following the ideas of Ref.~\onlinecite{Krylov.2013}. The batch is distributed such that a subset $\overline{ab}$ is chosen so that the batch fits into the available memory to ensure optimal usage of memory and to minimize overhead. Calculations are thus in principle only limited by the size of $OOVV$ quantities. The intermediate $W^{(t)}_{\overline{ab}ef}$ is then antisymmetrized and due to the fact that the full range of the index pair $ef$ is kept, the antisymmetry can be exploited in the subsequent contraction to the target amplitude:
\begin{equation}
 \begin{split}
     W^{(t)}_{\overline{ab}ef}&=\mathcal{P}^{-}_{\overline{ab}}\sum_J M_{\overline{a}e}^J M_{f\overline{b}}^{J*}\\
     t_{ij}^{\overline{ab}}&\leftarrow\sum_{e<f} W^{(t)}_{\overline{ab}ef}t_{ij}^{ef}\;.
 \end{split}
\end{equation}
Consequently, the intermediate $W^{(t)}_{abef}$ has to be calculated in every iteration which scales $V^4N_{\text{CH}}$. 
In practical applications this $N^5$ term though often scales worse than the particle-particle ladder term itself since usually $N_{\text{CH}}>O^2$ resulting in an overhead for the CD implementation in general. 
Still, compared to a canonical CCSD implementation, the advantage is that all tensors can be held in memory and the required intermediates are calculated as needed. Therefore, no disk-IO is needed to read in the two-electron-integrals $\bra{ab}\ket{cd}$ which poses to be a bottleneck of canonical CCSD.\newline
e) Concurrent use of single-precision apart from the CD can be exploited to further lower memory and computation-time requirements with virtually no loss in accuracy as shown for many post-HF methods such as CD-MP2\cite{Cederbaum.J.Chem.TheoryComput.2010} and CD-EOM-CCSD\cite{Krylov.J.Chem.TheoryComput.2018}. If higher accuracy than that provided when using single-precision is needed it is possible to {first converge the amplitudes to single-precision and then} switch to double precision {for the last iterations} to recover full accuracy.\cite{DePrinceJ.Chem.TheoryComput.2011,Krylov.J.Chem.TheoryComput.2018} 
%\textcolor{red}{As CC is solved using an iterative procedure this results in convergence to the correct double-precision amplitudes. This also holds in the ff-CD-CC context which was verified numerically.}
%

%%%%%%%%%%%%%%%%%%%%%%%%%%%%
\subsubsection{Validation}\label{sec:Validation}
%%%%%%%%%%%%%%%%%%%%%%%%%%%%
%
Validation is straightforward by comparing results with a tight Cholesky threshold to the canonical implementation employing the full ERI. 
Therefore, the test set used in Ref.~\onlinecite{Blaschke.J.Chem.Phys.2022} was employed.\footnote[4]{
The test set constitutes of the closed-shell molecules water and ethane, as well as the open-shell methylidyne radical in various magnetic field orientations and basis sets. 
Only the largest systems, i.e., CH using the unc-aug-cc-pV5Z basis and ethane using the unc-aug-cc-pVTZ basis set in a skewed orientation of the magnetic field were omitted. }
In Fig.~\ref{fig:plot_error}, the mean error of the ff-CD-HF energy, ff-CD-CCSD {and CC2} correlation energy and the ff-CD-EOM-CCSD {and CC2} excitation energy is plotted against the chosen Cholesky threshold. 
In accordance with previous results\cite{Blaschke.J.Chem.Phys.2022} not only the error in the integrals but also the error in the energies is bound by the chosen Cholesky threshold (black line). Up to a Cholesky parameter of $\delta=7$ the error of all methods is strictly below the chosen threshold. In this case, the error for (EOM)-CC is approximately one order of magnitude smaller than that of the HF energy.
Beyond this threshold the error is additionally constrained by the choice of the convergence criteria which is globally set to $10^{-7}$. Thus, the error in the EOM-CC energies becomes larger than the Cholesky threshold, while HF retains a good accuracy in the energy for orbital coefficients converged \textcolor{red}{to} $10^{-7}$.
\begin{figure}[tb]
    \centering
    \includegraphics[width=1.\linewidth]{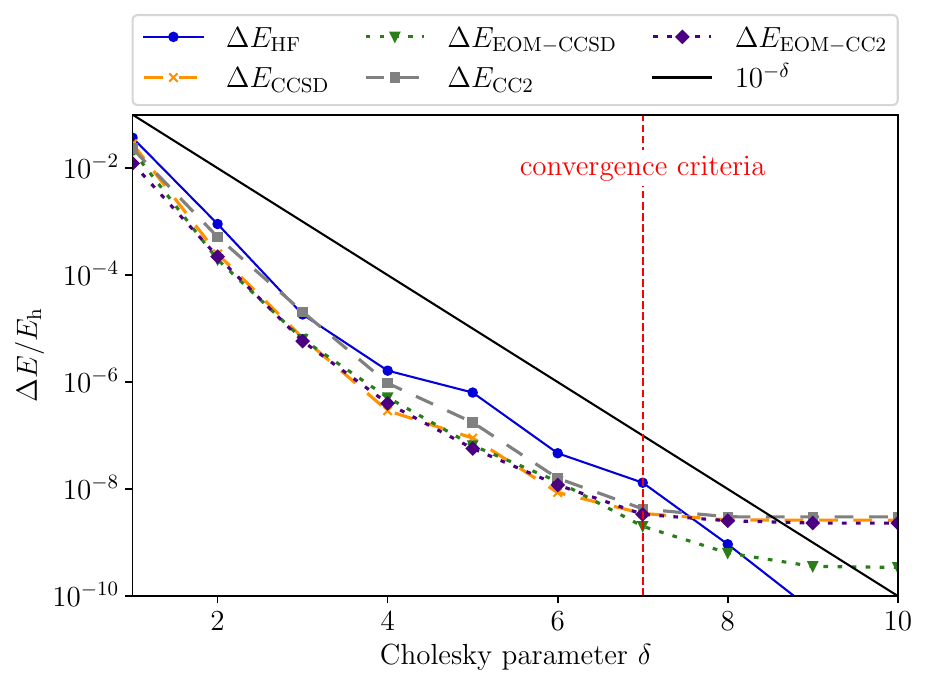}
    \caption{Mean error of the ff-CD-HF energy (blue), ff-CD-CCSD {and CC2} correlation energy (green {and grey}) and the ff-CD-EOM-CCSD {and CC2} excitation energy (orange {and indigo}) as a function of the Cholesky parameter $\delta$. The convergence criteria for HF, CC as well as EOM-CC were set to 10$^{-7}$.}
    \label{fig:plot_error}
\end{figure}

%%%%%%%%%%%%%%%%%%%%%%%%%%%%
\section{Applications}\label{sec:Application}
%%%%%%%%%%%%%%%%%%%%%%%%%%%%
%%%%%%%%%%%%%%%%%%%%%%%%%%%%
\subsection{Computational details}
%%%%%%%%%%%%%%%%%%%%%%%%%%%%
%
The calculations have been performed using two program packages. 
The implementation of ff-{CD-CC} was done within the QCUMBRE\cite{QCUMBRE} program package which handles the post-HF treatment. 
It is interfaced to CFOUR\cite{cfour,cfour.2020} with use of the MINT integral package.\cite{MINT} CFOUR handles the CD of the ERIs as well as the HF step. 
The following calculations on MgHe systems were performed using uncontracted (unc) augmented (aug)  correlation consistent basis sets. \cite{Dunning.cc,Dunning.aug}
The unc-aug-cc-pCVQZ\cite{Dunning.pCVXZ} basis was used for Mg and the unc-aug-cc-pVDZ basis was used for He if not stated otherwise {and all electrons were correlated}. 
The use of a larger basis set for helium produces only a small difference in the range of few percent, as validated in Tab.~S1 in the SI.
In the following we will discuss the $^3P\rightarrow {^3}S$ transition of triplet Mg mainly described by the excitation from 3p $\rightarrow$ 4s in different environments. 
Based on the symmetry of the studied system the respective states transform as  different irreducible representations which are listed in Tab.~\ref{tab:Symmetry_Table}. In all cases the states corresponding to the $^3P_{-1}$ state was chosen as EOM-CC reference.
\begin{table}[htb]
    \centering
    \caption{Symmetry labels for the relevant states for $3p$ to $4s$ transition of Mg in its triplet state considered in this study: The Mg atom ($SO(3)$), the Mg atom in a magnetic field ($C_{\infty h}$), MgHe ($C_{\infty v}$), MgHe in a parallel ($C_{\infty}$) and perpendicular field ($C_s$) as well as \ce{MgHe12} in a field aligned along the $C_3$ axis ($C_{3h}$)}
    \label{tab:Symmetry_Table}
    \begin{tabular*}{0.48\textwidth}{@{\extracolsep{\fill}}ll|lll|l}
        $SO(3)$             & $C_{\infty h}$             & $C_{\infty v}$           & $C_{\infty}$             &  $C_s$ &   $C_{3h}$\\ \midrule
      \multirow{3}{*}{$^3P$}& $^3\Sigma_u$               & $^3\Sigma^+$             & $^3\Sigma$               &  $A'$  &    $A'$\\
                            &  \multirow{2}{*}{$^3\Pi_u$}& \multirow{2}{*}{$^3\Pi$} & \multirow{2}{*}{$^3\Pi$} &  $A''$ &   \multirow{2}{*}{$E'$}  \\
                            &                            &                          &                          &  $A'$  &    \\ 
    \end{tabular*}
\end{table}

%%%%%%%%%%%%%%%%%%%%%%%%%%%%
\subsection{Pressure broadening of magnesium spectral lines}
%%%%%%%%%%%%%%%%%%%%%%%%%%%%

\begin{figure*}[t]
    \centering
    \includegraphics[width=1.\linewidth]{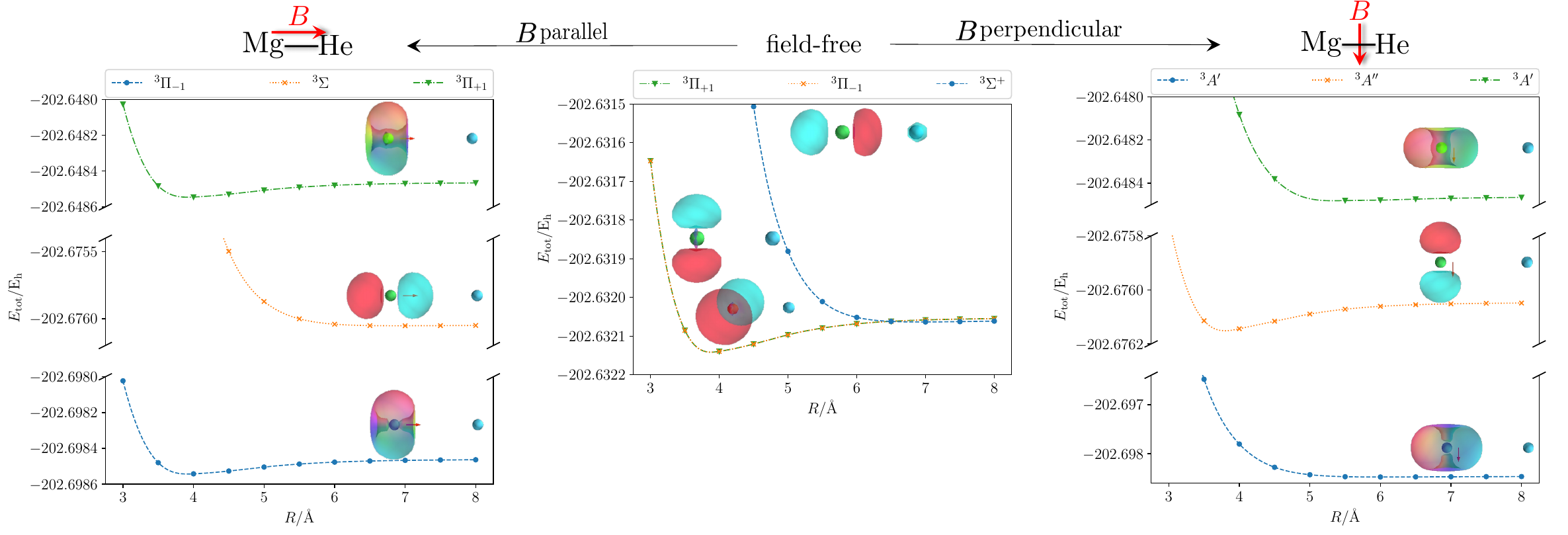}
    \caption{Total energy $E_\mathrm{tot}$ of the initial and first two final states of the MgHe triplet dimer without an external magnetic field (middle) and within a finite magnetic field of \SI{0.05}{\Bau} oriented parallel (left) and perpendicular (right) with respect to the Mg-He axis. The calculations were performed at the unc-aug-cc-pVTZ/CC3\cite{Kitsaras.J.Chem.Phys.2024} level of theory.\textcolor{blue}{$^a$} Additionally, the respective (complex) orbitals visualized via COrbit19\cite{Petros.2023.Dissertation} are shown. Mg is depicted in green and He in blue.\\[3pt]{\textcolor{blue}{$^a$} \footnotesize Note that an accuracy beyond CCSD is required to describe the dissociation limit of the degenerate $\Pi$ and $\Sigma$ states in a qualitatively correct manner.}}
    \label{fig:MgHe_CC3}
\end{figure*}

Assigning spectral lines for magnetic WDs is a non-trivial task. 
WDs are known to typically have dense helium or hydrogen atmospheres, and magnetic WDs in addition exhibit non-uniform surface-field distributions. Furthermore, even the coolest (magnetic) WDs have temperatures ranging from 4000 to 10000~K.\cite{Ferrario.Space.Sci.Rev.2015, Hollands2015, Hollands2017, Hollands2018, Hollands2023}
The 3p $\rightarrow$ 4s ($^3P\rightarrow {^3}S$) transition for the Mg triplet is one of the most intense lines for metal-containing WDs.\cite{Allard.Astron.Astrophys.2016,Hollands2023}
In Ref.~\onlinecite{Hollands2023}, this transition supported the assignment of a spectrum from a WD containing metals in its atmosphere (termed DZ) and a strong magnetic field of about \SI{3000}{\tesla}.
Due to the high surface gravities of WDs, one would expect the metals to sink down quickly, leaving pure He/H atmospheres. The fact that metals are nonetheless found is linked to accretion from planetary disks into the atmosphere.\cite{Allard.Astron.Astrophys.2016,Veras2021} 
This causes the star's atmosphere to be polluted by traces of metal species giving rise to distinct spectral lines. 
In order to study atoms and molecules under such extreme conditions it is crucial to model the surrounding environment, namely temperature, pressure, and density, as well as magnetic field, since they have a direct influence on the respective spectral lines.\newline
Variations in the surface magnetic-field strength and relative orientation of magnetic-field vector to the species of interest directly influence the transition wavelengths through alteration of the electronic structure. 
In the simplest model, to account for distribution of magnetic-field strengths over the star, a dipole is assumed, leading to a factor of two over the encountered field strengths.
Due to their variation and due to the fact that the magnetic field strengths are not a priori known, field-dependent  transition wavelengths ($B - \lambda$ curves) are required to help assignment of spectra.\newline
High-density atmospheres, where collisions are frequent, are prone to exhibit strong pressure-broadening effects. 
Especially due to the relatively low opacity of helium atmospheres at lower effective temperatures, observed photons can originate from deeper atmospheric layers with higher densities.\cite{Bergeron1995, Allard.Astron.Astrophys.2016, Blouin2020}
Collisions between metal and surrounding atmosphere atoms alter the energy levels of the radiating metal atom and thus cause a shift as well as a broadening of the spectral lines in the WD's spectrum. 
The broadening arises from the fact that the distance between the radiating atom and atmosphere particles is not constant. 
Instead, it is distributed around a mean value dependent on pressure and temperature. 
Moreover, the {individual} energy levels involved in a transition  are affected due to the interaction with the environment.\cite{LaserSpectroscopy.Demtroeder2002}
{The pressure effects cause} both the shift as well as asymmetric line shapes and satellites as discussed, for example, in Ref.~\onlinecite{Allard.Astron.Astrophys.2016}. For a review on methods to model the spectral lines see also Ref.~\onlinecite{Allard1982}. Generally, a parcel of gas at a specific temperature and number density of helium atoms is assumed. Within that parcel some helium atoms will collide with a radiator atom and consequentially the energy levels and transition dipole-moments of the radiator will be altered strongly. Statistically, other helium atoms will be further away and therefore will not significantly perturb the system. To calculate the line profile for this parcel of gas, integration over the distribution of atomic separations is performed. Consequently, only the diatomic potentials are needed to model the line shape.\footnote[5]{The method considers the simultaneous interaction with multiple perturbers but neglects the interperturber correlation, i.e., the interaction between the perturbers.\cite{Allard2023}}\newline
{\subsubsection{MgHe interaction}}
The polarization by surrounding helium atoms can significantly affect the electronic structure of magnesium and thus warrants investigation in order to be able to model the spectral line position and shape. In this study, we investigate the pressure effects on the 3p $\rightarrow$ 4s ($^3P\rightarrow {^3}S$) transition for the Mg triplet perturbed by a dense helium atmosphere in an external magnetic field. A number of previous studies\cite{Allard1982, Allard2014, Allard.Astron.Astrophys.2016, Allard2016, Allard.Astron.Astrophys.2018, Blouin2018, Allard2023} have already been conducted in the field-free case, where Refs.~\onlinecite{Allard.Astron.Astrophys.2016} and \onlinecite{Hollands2017} specifically deal with the Mg triplet transition of interest. Here, we will build upon the results of previous studies by expanding the discussion by the influence of an external magnetic field on the pressure effects:\newline
We start our discussion on the Mg-He dimer in the field-free case. Due to the fact that the asymmetric broadening can be observed whether or not spin-orbit splittings are resolved within it,\cite{Allard.Astron.Astrophys.2016,Hollands2017} we neglect their contribution for our discussion.\footnote[6]{The size of the spin-orbit contribution amounts to approximately \SIrange{5}{10}{\angstrom}.\cite{NIST_ASD}}
Starting from the field-free case, the initial state for the isolated Mg atom in the $^3P$-state is described by the three degenerate $p$-orbitals. 
Introducing the helium atom into the system lowers the molecular symmetry to $C_{\infty v}$. 
As a result the former $^3P$-state splits into two degenerate bonding $^3\Pi$-states whose highest occupied molecular orbitals (HOMOs) are perpendicular to the bond axis. 
The $^3\Sigma^+$(3p$_0$) state, with the HOMO oriented along the bond axis, is non bonding and higher in energy as shown in the central panel of Fig.~\ref{fig:MgHe_CC3}. 
The transition wavelength of the Mg transition ($^3P\rightarrow{^3}S$) as a function of the distance to the helium atom is shown in Fig.~\ref{fig:Shift_vs_R} together with the total energies of the involved $^3\Pi$ and $^3\Sigma^+$(4s) states.
Note that the unperturbed transition wavelength in our calculation is about \SI{5132}{\angstrom} while the observed one is centred around \SI{5174}{\angstrom}.\cite{NIST_ASD} 
{
The difference is due to basis-set incompleteness errors, higher-order correlation effects, as well as relativistic effects.}
This discrepancy does {not}, however, impact the overall qualitative analysis as the focus is on energy shifts which are not expected to change significantly. Still, as already discussed in Ref.~\onlinecite{HampePhys.Chem.Chem.Phys.2020}, for fully quantitative predictions, composite schemes would need to be employed. {These may include, as mentioned above, an extrapolation to the basis-set limit, higher-order correlation effects,  and a shift to the NIST data to account for field-free scalar relativistic effects.}\newline
As the helium atom approaches an isolated magnesium atom, a blue shift arises in a gradual manner. 
Until a distance of \SI{9.0}{\angstrom}, the interaction is so weak that the change in the transition wavelength is smaller than \SI{1}{\angstrom}. 
Thus, up to this distance the system can be treated as non-interacting. 
The blue shift increases with the shortening of the Mg-He distance and reaches its maximum with a shift of about \SI{108}{\angstrom} at \SI{3.5}{\angstrom}. 
The main reason for the shift is the occurrence of a shoulder for smaller distances in the final $^3\Sigma^+$ state.
For shorter distances the transition wavelength rises up to an inter-atomic separation of \SI{2.0}{\angstrom}. 
While the initial state becomes more and more repulsive, the final state stays relatively unchanged, leading to a gradually increasing red shift relative to the unperturbed case. 
\newline
Applying an external magnetic field changes the interaction in the MgHe dimer as shown in Fig.~\ref{fig:MgHe_CC3}. 
\begin{figure}[tb]
    \centering
    \includegraphics[width=1.\linewidth]{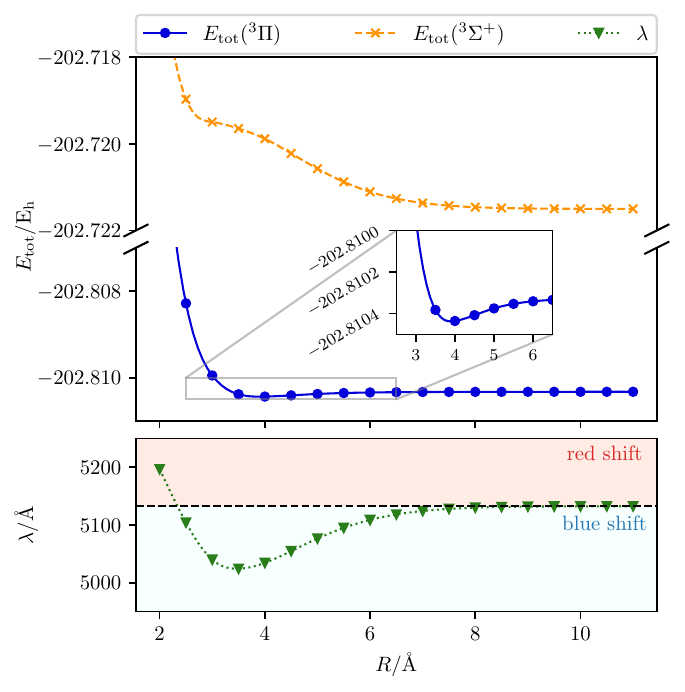}
    \caption{The transition wavelength $\lambda$ ($^3\Pi\rightarrow{^3}\Sigma^+$) as well as the total initial-state energy $E_{\mathrm{tot}}$ as a function of the Mg-He distance $R$. Calculated with CD-CCSD ($\delta=5$) using the unc-aug-cc-pCVQZ basis for Mg and the unc-aug-cc-pVDZ basis for He.}
    \label{fig:Shift_vs_R}
\end{figure}
In principle, all orientations of the magnetic field with respect to the bond axis should be taken into account. Here, we are picking out the two extreme cases, i.e., the parallel and the perpendicular orientations. 
In a magnetic field, the individual states are altered in the following ways: In the parallel orientation ($C_{\infty}$ symmetry), the bonding $^3\Pi_{-1}$ state is stabilized and becomes the new ground state while the $^3\Pi_{+1}$ state is destabilized. The non-bonding $^3\Sigma$-state has no orbital-Zeeman contribution and constitutes the first excited state. Conversely, in the perpendicular orientation ($C_s$ symmetry) the non-bonding { $^3A'$}-state is stabilised and becomes the ground state while the first excited { $^3A''$}-state is bonding. 
Increasing the magnetic field strength further, up to \SI{0.20}{\Bau}, results in a slight destabilisation of the bound ground state in the parallel orientation. In the perpendicular orientation, when increasing the magnetic-field strength, the ground state becomes increasingly bound via the orbital-Zeeman interaction, as can be seen in Fig.~\ref{fig:pothyp}. 
\begin{figure*}[tb]
     \centering
     \begin{subfigure}[b]{0.49\linewidth}
         \centering
         \includegraphics[trim={0.75cm 0 0 0},clip,width=\linewidth]{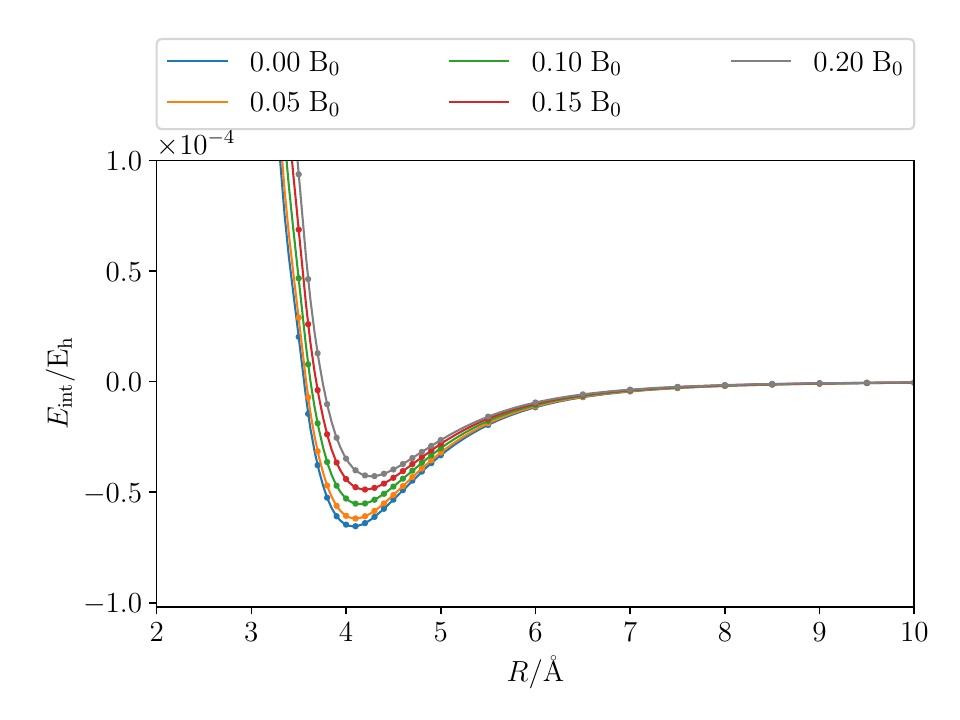}
         \caption{$^3\Pi_{-1}$ (parallel field)}
         \label{fig:pothyp_parallel}
     \end{subfigure}
     %\hfill
     \begin{subfigure}[b]{0.49\linewidth}
        \centering
         \includegraphics[trim={0.75cm 0 0 0},clip,width=\linewidth]{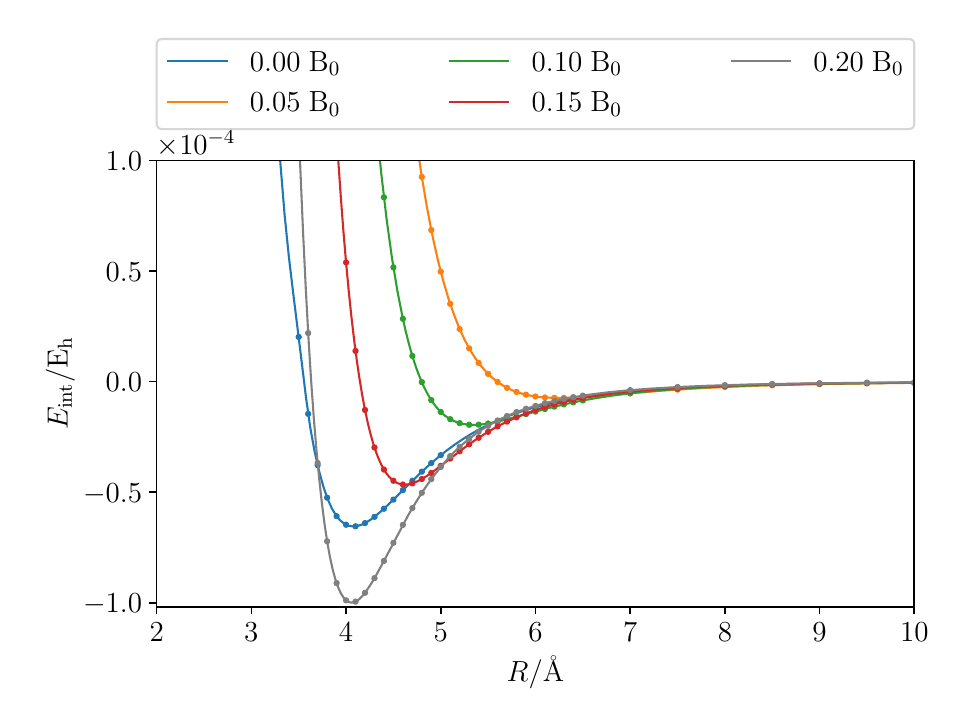}
         \caption{$^3A'$(perpendicular field)}
         \label{fig:pothyp_perpendicular}
     \end{subfigure}
     %\hfill
    \caption{The counterpoise corrected\cite{Boys.1970,vanDuijneveldt1994} CCSD interaction energy of the initial state of the MgHe triplet dimer plotted as a function of the external magnetic field strength in a parallel (\ref{fig:pothyp_parallel}) and perpendicular (\ref{fig:pothyp_perpendicular}) orientation, respectively. The calculations were performed using the unc-aug-cc-pCVQZ basis for Mg and an unc-aug-cc-pVDZ basis for He.}
    \label{fig:pothyp}
\end{figure*}

In order to assess the influence of the perturbing helium atom on the spectral lines of the magnesium triplet within a magnetic field the corresponding  $B-\lambda$ curves, i.e., the magnetic field strength plotted against the transition wavelength, are investigated as shown in Fig.~\ref{fig:Blambda}.
Note that we follow here the main character of the transition for isolated Mg such that the corresponding $B-\lambda$ curves remain comparable even though formally there are a couple of avoided crossings with other states involved, particularly for higher magnetic-field strengths.
For further details we refer to Figs.~S1 and S2 in the SI.
For the calculation, the respective equilibrium Mg-He distance within a given magnetic-field strength and orientation was chosen (see Tab.~S13 in the SI). 
\begin{figure}[tb]
    \centering
    \begin{overpic}[width=1.\linewidth]{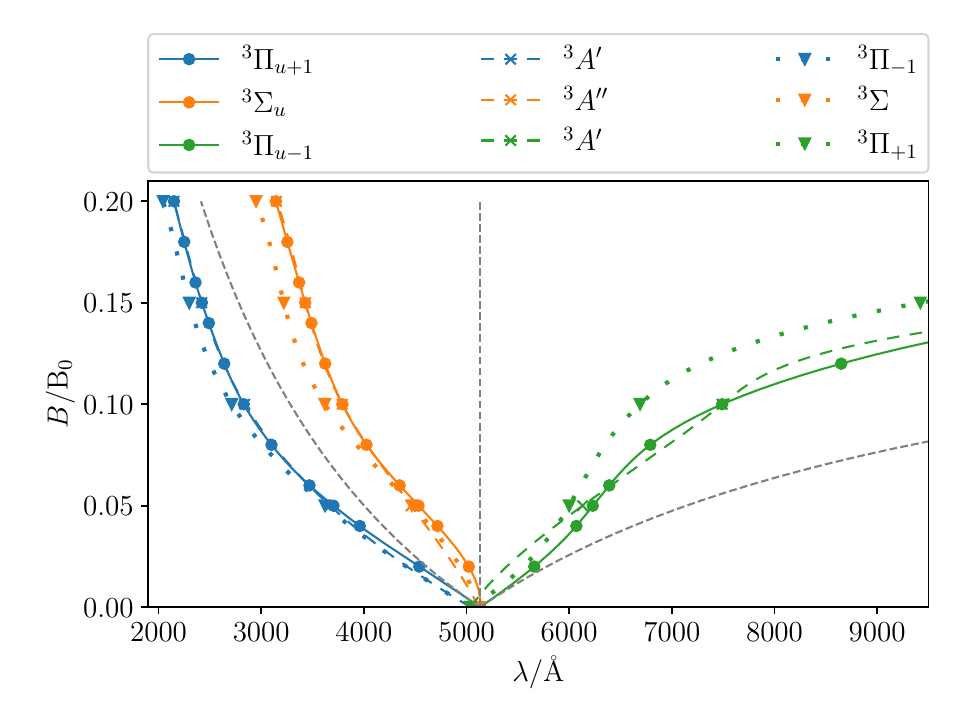}
    \put(43,185){\centering Mg}
    \put(95,185){\centering MgHe perpendicular}
    \put(188,185){\centering MgHe parallel}
    \end{overpic}
    \caption{$B-\lambda$ curves for the transitions from the three former $P$-states. Three cases are distinguished. The isolated Mg atom is used as a reference ($C_{\infty h}$, solid lines) and is compared with the MgHe triplet dimer at the equilibrium geometry in the parallel ($C_{\infty}$, dotted lines) or the perpendicular ($C_s$, dashed lines) magnetic field. Calculated at the CD-CCSD ($\delta=5$) level using the unc-aug-cc-pCVQZ basis for Mg and an unc-aug-cc-pVDZ basis for He. Additionally, the simple orbital-Zeeman split (dashed gray lines) is also shown.}
    \label{fig:Blambda}
\end{figure}
The $p$-orbitals, which are degenerate in the field-free case, split into three components resulting in large shifts in the transition wavelength.
Additionally, due to the interplay of para- and diamagnetic effects induced by the external magnetic field, the lines vary in a non-trivial manner as compared to a simple orbital-Zeeman shift. 
Hence, ff methods are indispensable for interpreting spectra of strongly magnetic WDs as discussed in detail in Ref.~\onlinecite{Hollands2023}.\textcolor{blue}{\footnote[5]{Note also that the transition-dipole moments change as a function of the magnetic field and the orientation of the magnetic field as can be seen in Fig.~S5 in the SI.}}\newline
To assess the influence of the Helium atom on the transition in the magnetic field in more detail, the shift defined as the difference between the excitation energy of the isolated magnesium atom and the dimer in a parallel or perpendicular orientation is shown in Fig.~\ref{fig:Shift_vs_B}.
\begin{figure}[tb]
    \centering
    \begin{overpic}[width=1.\linewidth]{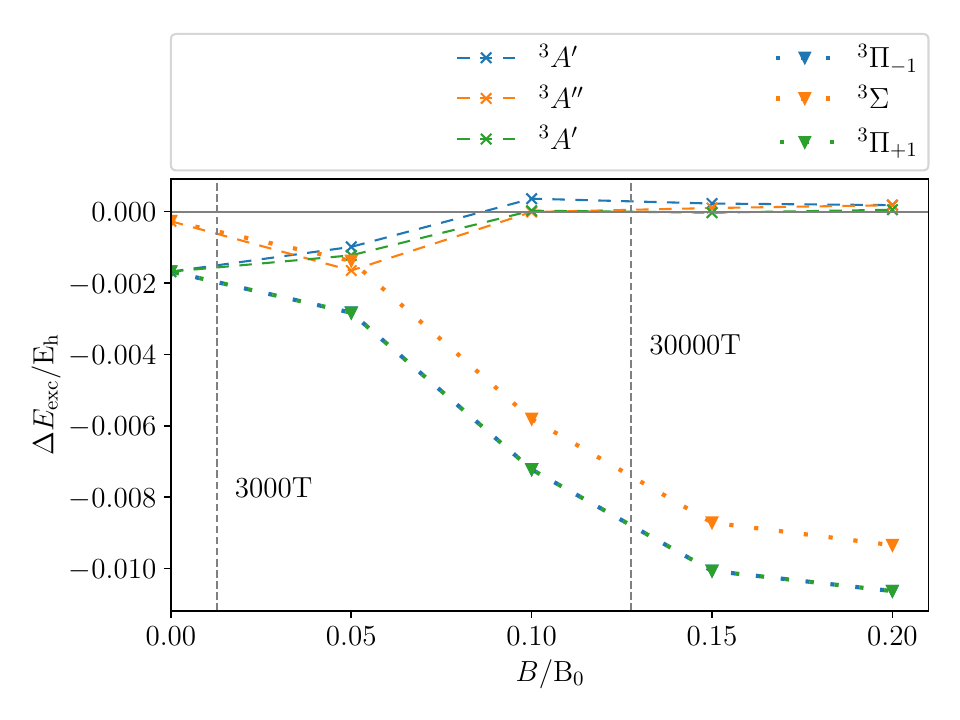}
    \put(95,185){\centering MgHe perpendicular}
    \put(188,185){\centering MgHe parallel}
    \end{overpic}
    \caption{Shift in the excitation energies $\Delta E_{\mathrm{exc}}$ induced by the helium atom for the three former $P$-states of the MgHe triplet dimer at the equilibrium geometry in a parallel ($C_{\infty}$, dotted) or perpendicular ($C_{s}$, dashed) magnetic field compared to the isolated Mg atom. For the changes in the transition wavelengths see also Fig.~\ref{fig:Blambda}.} 
    \label{fig:Shift_vs_B}
\end{figure}
In the magnetic field, the following can be observed:
\begin{enumerate}
\item
The parallel orientation leads to very large blue shifts that increase with the magnetic field. This is explained by the fact that the $^3\Sigma$(4s) final state is destabilized\footnote[3]{The $^3\Sigma^+$ state acquires d-character via an avoided crossing as discussed in Ref.~\onlinecite{Hollands2023}. 
This polarization is apparent in the shape of the HOMO (4s in the field-free case) which acquires d$_0$ orbital-like lobes as shown in Fig.~S3 in the SI. In the parallel orientation, this leads to an antibonding and hence destabilizing interaction with He (see also Figs.~S1 and S2 for the total energies)} in the magnetic field as compared to the atomic case while the initial states remain mostly unchanged. In addition, it develops an increasingly large shoulder (see also Figs.S6 and S7 in the SI) which leads to larger shifts. 
\item
Furthermore, in the parallel orientation, the curves for the shift in the transitions from the $^3\Pi$ and $^3\Sigma$ states, respectively, remain parallel. 
In the absence of a magnetic field, the influence of the Helium atom is stronger for transitions from the bonding $^3\Pi$ states than for the non-bonding $^3\Sigma^+$(3p$_0$) state. 
The difference is due to the fact that the excitation energy is evaluated at the equilibrium distance of the $^3\Pi$ state at which the $^3\Sigma^+$ (3p$_0$) is repulsive. 
This in turn compensates to some degree the shoulder in the final $^3\Sigma^+$(4s) state (see Fig.~\ref{fig:Shift_vs_R}). 
Thus, compared to the isolated Mg atom the transition from the bonding $\Pi$ states is blue shifted as shown in Fig.~\ref{fig:Shift_vs_R} while the transition from $^3\Sigma^+$ (3p$_0$) is relatively unaffected. 
As the initial states remain mostly unchanged in the parallel field, so does their respective relative position. 
As such the shift for the transitions from the $^3\Sigma^+$(3p$_0$) state evolves in parallel to those from the $^3\Pi$ states when increasing the field strength.
\item 
For the perpendicular orientation, the shifts vanish for large magnetic-field strengths. 
In this orientation, the states evolve in a comparable manner to the atomic case (see Fig.~S1 in the SI) and the shoulder in the final state diminishes (see also Figs.~S6 and S7 in the SI). \footnote[4]{This is due to the fact that the polarization of the 4s orbital is perpendicular to the Mg-He bond which leads to more atomic-like states, see Fig.~S3 in the SI.}  
\end{enumerate}
For the strongest magnetic field considered (\SI{0.20}{\Bau}) the shift due to the magnetic field in the perpendicular orientation is of the order of \SI{0.2}{\milli\Hartree}. In the parallel orientation shifts of about \SI{10.6}{\milli\Hartree} arise. These cases correspond to shifts in the wavelength of about $\SI{1}{\angstrom}$ and $\SI{100}{\angstrom}$, respectively.\newline
{\subsubsection{MgHe cluster}}
The spectrum of a real white dwarf atmosphere is a result of the light emitted from various atmospheric layers including deeper layers which have significantly larger densities than the higher layers. It is quite typical for atmospheric layers that contribute to the observable spectra of WDs to reach densities of $n(\mathrm{He})\approx 10^{22}\text{ to }10^{23}\;\SI{}{\per\centi\metre\cubed}$, where the dense helium atmosphere shows fluid-like behaviour.\cite{Bergeron1995, Kowalski2004, Kowalski2006, Kowalski2010, Allard.Astron.Astrophys.2018} 
For sufficiently large densities the average distance between the helium atoms and the magnesium is smaller than the collision cross section. Under such conditions, the radiating magnesium atom experiences numerous concurrent perturbations as it interacts with the adjacent helium atoms.\newline 
\begin{figure}[tb]%
    \centering
    \includegraphics[width=1.\linewidth]{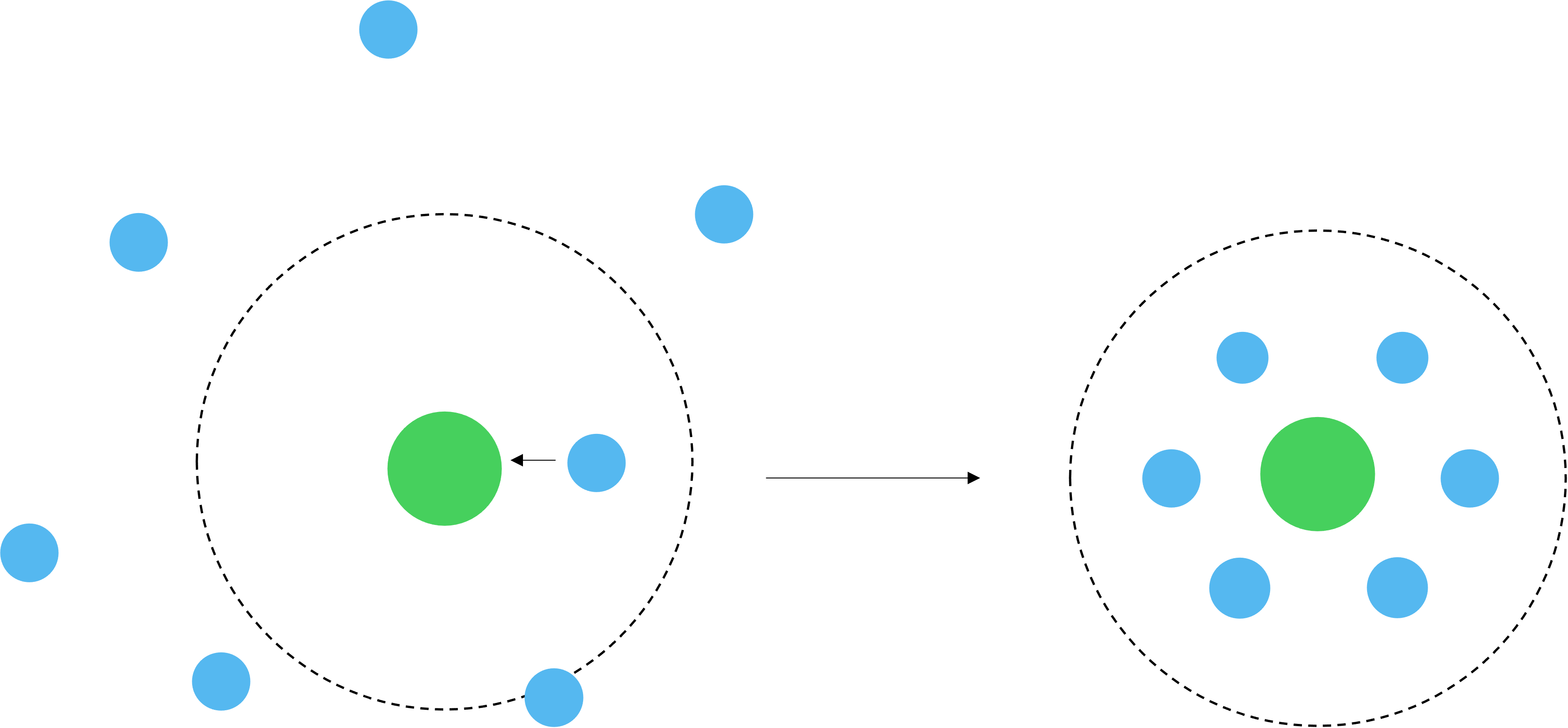}
    \caption{Schematic depiction of the transition between the MgHe dimer model system to an explicit solvation model for large densities. The collision cross section is shown as dotted circle (Mg: green, He: blue).}
    \label{fig:Atmossphere}
\end{figure}%
To account for such a perturbation, we also investigate Mg in an explicit solvation model (see Fig.~\ref{fig:Atmossphere}).
A magnesium atom is placed in an idealized hexagonal closed packed structure, where all atoms within the cluster have the same interatomic distance. This may be a reasonable choice of configuration since helium crystallizes in a hcp structure under high pressures.\footnote[7]{Note also that the face centered cubic (fcc) lattice has the same density and thus is expected to show similar pressure effects.} The hcp structure with a given radius can then be connected to a density of a helium atmosphere.\newline
In order to ensure that the treatment of the first coordination sphere (\ce{MgHe12}) is sufficient to model the influence of the helium atmosphere in an explicit solvent model the contribution of the second shell is evaluated using the \ce{MgHe56} cluster depicted in Fig.~\ref{fig:MgHe56} (using a lattice constant of \SI{3}{\angstrom} which corresponds to a density of \SI{6.5E21}{\text{atoms}\per\centi\metre\cubed})\footnote[8]{Note that the distance between two atoms is twice the lattice constant.}. Therefore, a CD-CC2 calculation using the unc-aug-cc-pCVQZ basis on Mg and the unc-aug-cc-pVDZ basis on He was performed. 
We note that in this system with 781 basis functions the memory demand to store the full integrals in AO basis would correspond to almost 60~TB. 
This prohibitively large memory demand is diminished by using CD, which generates 3505 vectors ($\delta=5$) with a memory requirement of 34~GB. Overall, CD is  mandatory in order to make a computational description of such systems  at the ff-CC level feasible.\newline 
{Concerning the accuracy of ff-CC2, we note that as long as the states involved do not have a strong double-excitation character, the development of the excitation energies as a function of the magnetic field is reproduced faithfully. 
For MgHe$_12$, the deviations between ff-EOM-CC2 and ff-EOM-CCSD, are around 1mEh (see Tab. S11 and S12 in the SI).
For more details on the performance of ff-CC2, we refer the reader to Ref. \onlinecite{Kitsaras.J.Chem.Phys.2024}}.
Compared to \ce{MgHe12} the second layer contributes only \SI{7}{\angstrom} to the shift (see SI Tab.~S12). This is expected as a) the polarization of the outer atoms is shielded by the first shell and b) as shown in Fig.~\ref{fig:Shift_vs_R} the shift decreases rapidly with the distance to the Mg atom. This validates our approach to use the first coordination sphere as model system.\newline
\begin{figure}[tb]%
    \centering
    \includegraphics[width=0.8\linewidth]{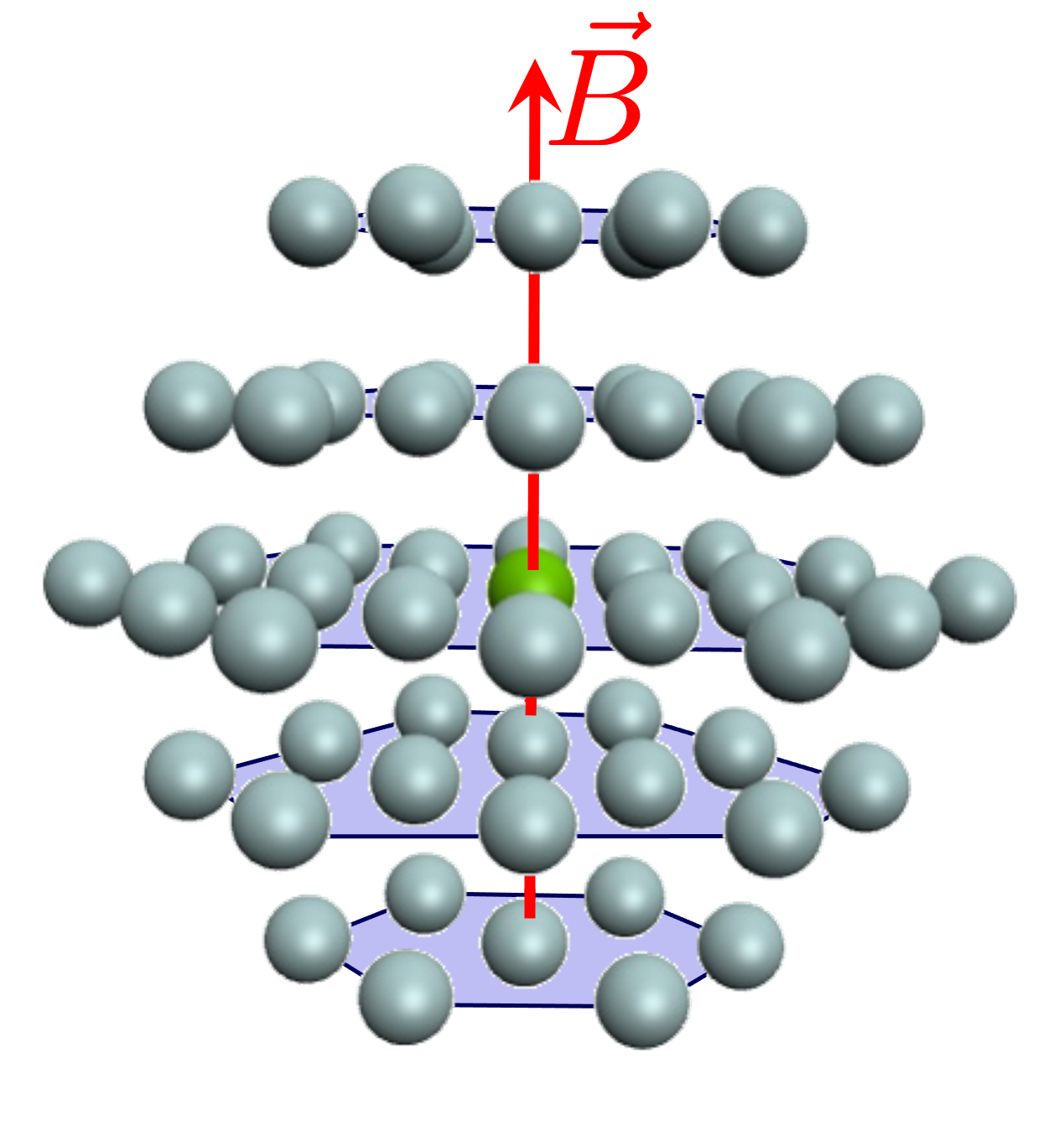}
    \caption{Representation of a Mg atom (green) in a hcp lattice of He atoms (gray). Depicted are the first two shells forming a cluster of \ce{MgHe56}, where 12 He atoms are in the first and 44 He atoms in the second shell.}
    \label{fig:MgHe56}
\end{figure}%
\begin{figure}[tb]%
    \centering
    \begin{overpic}[width=1.\linewidth]{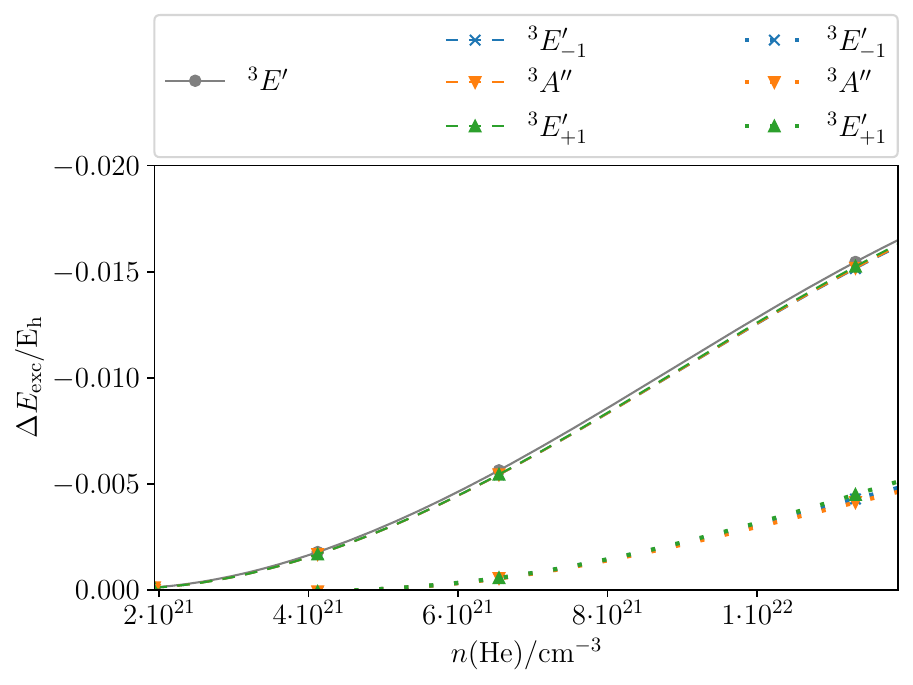}
    \put(44,188){\centering  B=\SI{0.0}{\tesla}}
    \put(121,188){\centering B=\SI{3000}{\tesla}}
    \put(208,188){\centering  B=\SI{30000}{\tesla}}
    \end{overpic}
    \caption{Shift in the excitation energy $\Delta E_{\mathrm{exc}} = E_{\mathrm{exc}}(\rho\rightarrow0)-E_{\mathrm{exc}}(\rho)$ (in Hartree) at the CD-CCSD ($\delta=5$) level of the Mg triplet transition as a function of the helium density $\rho$ of the \ce{MgHe12} ($C_{3h}$) atmospheric model system in the field-free case as well as in a magnetic field of \SI{3000}{\tesla} and \SI{30000}{\tesla} oriented along the $C_3$ axis. Employing the unc-aug-cc-pCVQZ on Mg and the unc-aug-cc-pVDZ basis on He.}
    \label{fig:MgHe12_shift_vs_density_hartree}
\end{figure}%
Fig.~\ref{fig:MgHe12_shift_vs_density_hartree} shows the shift in the excitation energies for the \ce{MgHe12} cluster of the first coordination sphere oriented along the $C_3$ axis as a function of helium density. The field-free case is shown in grey. 
Similar trends as for the MgHe dimer can be observed: 
For densities lower than \SI{1E21}{\text{atoms}\per\centi\metre\cubed} the excitation energy and thereby also the spectral line is not shifted as the distance between Mg and He is too large. 
For denser atmospheres a blue shift, as expected from Fig.~\ref{fig:Shift_vs_R}, arises which gets stronger with decreasing interatomic distance, i.e., increasing the density. 
Thereby, the transition wavelength shifts as a function of the density by several hundred \SI{}{\angstrom} (see also Fig.~S4 in the SI) which corresponds to ten to thirteen times the effect of a single helium atom on the transition wavelength. 
However, the measured signals are emitted from various atmospheric layers, and hence from a distribution of temperatures and densities. 
Deeper layers with large densities can be expected to have a larger Rosseland mean opacity.\cite{Rosseland1924} {The Rosseland mean opacity is a weighted mean of the total opacity over all wavelengths, which, among other factors, depends on the density, temperature, and composition (mass fractions) of the star. 
For our purposes, it can be understood as a measure of the number of photons that exit the stellar atmosphere.}\footnote[9]{{Note that the opacity is connected to the Rosseland optical depth $\tau_R$ which describes the decay of the intensity of radiation $I=I_0e^{-\tau_{R}}$. For $\tau_R = 2/3$ the intensity is approximately halved which means half the flux radiated from that layer will undergo additional absorption and scattering but half will escape into space where it can be detected. }}\cite{Carroll2017}
Flux emitted from the deep, high-density layers of the atmosphere will undergo additional absorption and scattering, and only a fraction can be detected. 
Conversely, higher layers show an increased transparency where most of the flux is escaping the atmosphere and hence contribute to the observed spectrum.
Additionally, the transition-dipole moment and hence the respective oscillator strength for this transition decreases as well with increasing density (see SI Tab.~S11).
Consequently, the most intense signals originate from layers with moderate densities. 
The resulting overall line profile is thus not necessarily shifted by these large wavelength, but asymmetrically broadened.\cite{Allard.Astron.Astrophys.2016,Hollands2017} 
\newline
The dashed lines \textcolor{red}{in Fig.~\ref{fig:MgHe12_shift_vs_density_hartree}} show the pressure dependence for a magnetic field of \SI{3000}{\tesla}. 
The three components show a very similar behavior as a function of the density as in the field-free case. 
For moderate field strengths, such as \SI{3000}{\tesla}, it thus seems justifiable to simply include the atmospheric pressure effects of the field-free case as a model for the line shapes in the magnetic field.\newline
For stronger magnetic field strengths of \SI{30000}{\tesla} (dotted lines) the behavior is similar for the three components but it is rather different from the cases discussed before. 
In fact, the shifts are much smaller and become relevant only for very large densities. 
The overall lower shift in the excitation energy may be explained in the following manner: as discussed previously for Fig.~\ref{fig:Shift_vs_B}, an increasing vs. decreasing shift was found for the parallel and perpendicular orientations, respectively. 
For the \ce{MgHe12} cluster a superposition of these effect leads to an overall decrease of pressure effects for \SI{30000}{\tesla}. 
For strong magnetic fields it may hence be reasonable to disregard the pressure effects except for very large densities.\newline
Note however that because of the non-linear scale, if the shift is expressed in transition wavelengths (see Fig.~S4 in the SI), the differences in the shift are more pronounced. In particular, the $^3E'_{+1}$ component leads to much larger shifts than the other components. 
Again, for a quantitative prediction, composite schemes would need to be employed.\newline
In this first-order approach in which we are using a static model system, insights into the influence of the atmosphere on the Mg transitions are gained. 
Going a step further could imply using a QM/MM approach for a dynamic description which can capture the effects of an arbitrary magnetic-field orientation at a given temperature and pressure while retaining the needed high level accuracy at the radiating Mg atom which will be the subject of further studies.
%
%%%%%%%%%%%%%%%%%%%%%%%%%%%%
\section{Conclusion}\label{sec:conclusion}
%%%%%%%%%%%%%%%%%%%%%%%%%%%%
%
In this work we introduced the implementation of ff ground and excited-state CC methods exploiting the CD representation of the ERI (ff-CD-CC). 
The ff-CD-CC scheme allows to perform highly accurate calculations for atoms and molecules subjected to an external magnetic fields as can be found on magnetic WDs. 
By significantly reducing memory requirements, the CD tackles computational bottlenecks, enabling the investigation of systems which were previously computationally not feasible. \\
Assigning spectra from magnetic white dwarf stars requires very accurate predictions. 
It is a non-trivial task as such stars exhibit dense atmospheres, non-uniform strong magnetic fields, and a wide temperature range. 
These factors directly influence the spectral line shapes of atoms and molecules, making accurate modeling challenging.
We showed that these developments enabled the study of pressure broadening of magnesium spectral lines on magnetic WDs. 
Furthermore, we discussed the influence of an external magnetic field and the considerable differences between the parallel and perpendicular orientations. 
In the MgHe system, the change in excitation energies with respect to the magnetic-field strength shows that the pressure shift due to He increases in parallel and decreases in perpendicular orientations.
Lastly, we employed a cluster model to consider a more realistic scenario for high-density helium atmospheres.
Particularly for treating the cluster, CD techniques becomes crucial. 
The analysis showed that for \SI{3000}{\tesla} pressure effects in the field-free case are comparable to the effects in magnetic fields which can justify incorporating pressure effects from the field-free case as approximation for line shapes in magnetic fields.\newline
However, this approximation may not hold for stronger magnetic fields of, for example, \SI{30000}{\tesla} where the magnitude of the pressure effect on the excitation energies is decreased drastically.\newline
Overall, this study shows the complex interplay between magnetic fields and pressure effects, and their influence on the spectral lines of magnesium. Further investigations, including dynamic descriptions of the environment and inclusion of temperature effects, are necessary to gain a conclusive understanding of matter in magnetic WDs. 
%%%%%%%%%%%%%%%%%%%%%%%%%%%%%%%%%%%%%%%%%%%%%%%%%%%%%%%%%%%%%%%%%%%%%%%%%%%%%%%%%%%%%%%%%%%%%%%
\section{Conflicts of interest}
The authors have no conflicts to disclose.
\section{Data availability statement}
The data that supports the findings of this study are available within the article and its {supplementary material}.

\section{Acknowledgments}
The authors thank Jürgen Gauss, Filippo Lipparini, and Mark Hollands for helpful discussions. This work has been supported by the Deutsche Forschungsgemeinschaft (DFG) via project B05 of the TRR 146 “Multiscale Simulation Methods for Soft Matter Systems” and via Grant No. STO 1239/1-1.
\section{Supplementary material}
See \hyperlink{si}{supplementary material} for the details of the reported calculations. They consist of the raw data of the plots presented in the manuscript as well as additional plots and figures.
\section{ORCID}%
Simon Blaschke http://orcid.org/0000-0001-5033-228X \\
Marios-Petros Kitsaras https://orcid.org/0000-0002- 9549-3674 \\
Stella Stopkowicz https://orcid.org/0000-0002-0037-7962
%%%%%%%%%%%%%%%%%%%%%%%%%%%%%%%%%%%
\appendix
%%%%%%%%%%%%%%%%%%%%%%%%%%%%%%%%%%%
\section{Diagrammatic Approach}
%%%%%%%%%%%%%%%%%%%%%%%%%%%%%%%%%%%
%
In the diagrammatic sense, let a pair of Cholesky vectors be described by one-electron operators.
Thus, a vector and its complex conjugate can be defined as:
\begin{figure}[H]
    \centering
    \resizebox{\linewidth}{!}{
    \begin{tikzpicture}%[every path/.append style={color=mycyan}, every node/.append style={color=mycyan}]
        \startnewset<L>[$\begin{aligned}L_{ia}^J\end{aligned}$]{\diagram{   
            \OPo*[r](0,0)<L>
            \LINEVdr(L)[left]<a><i>
        }}
        \nextterm[][$\begin{aligned}L_{ai}^J\end{aligned}$]{\diagram{
            \OPo*[r](0,0)<L>
            \LINEVur(L)[left]<i><a>
        }}
        \nextterm[][$\begin{aligned}L_{ab}^J\end{aligned}$]{\diagram{
            \OPo*[r](0,0)<L>
            \LINEuul(L)<b><a>
        }}
        \nextterm[][$\begin{aligned}L_{ij}^J\end{aligned}$]{\diagram{
            \OPo*[r](0,0)<L>
            \LINEddl(L)<j><i>
        }}
    \end{tikzpicture} 
    }
\end{figure}
\begin{figure}[H]
    \centering
    \resizebox{\linewidth}{!}{
    \begin{tikzpicture}
        \startnewset<L>[$\begin{aligned}L_{ai}^{J*}\end{aligned}$]{\diagram{   
            \OPo*(0,0)<L>
            \LINEVdr(L)[left]<a><i>
        }}
        \nextterm[][$\begin{aligned}L_{ia}^{J*}\end{aligned}$]{\diagram{
            \OPo*(0,0)<L>
            \LINEVur(L)[left]<i><a>
        }}
        \nextterm[][$\begin{aligned}L_{ab}^{J*}\end{aligned}$]{\diagram{
            \OPo*(0,0)<L>
            \LINEuur(L)<a><b>
        }}
        \nextterm[][$\begin{aligned}L_{ij}^{J*}\end{aligned}$]{\diagram{
            \OPo*(0,0)<L>
            \LINEddr(L)<i><j>
        }}
    \end{tikzpicture} 
    }
\end{figure}
hence any (non-antisymmetrized) integral e.g $\braket{ij|ab}$ can be represented as
\begin{figure}[H]
    \centering
    \resizebox{0.75\linewidth}{!}{
    \begin{tikzpicture}
        \startnewset<L>[$\begin{aligned}\braket{ij|ab}\end{aligned}$]{\diagram{   
            \OPwgl*(0,0)<w>
            \LINEVdr(w-1)[left]<a><i>
            \LINEVdr(w-2)[left]<b><j>
        }}
        \nextterm[$=$][$\begin{aligned}\sum_J L_{ia}^{J} L_{bj}^{J*}\end{aligned}$]{\diagram{
            \OPo*[r](0,0)<L>
            \LINEVdr(L)[left]<a><i>
            \OPo*(1.5*\dvdisp,0)<L>
            \LINEVdr(L)[left]<b><j>
        }}
    \end{tikzpicture} 
    }
\end{figure}
where the left vector is read as $L_{out,in}^J$ and the right vector is defined as $L_{in,out}^{J*}$. Thus, the intermediates can be understood as the contractions
\begin{figure}[H]
    \centering
    \resizebox{\linewidth}{!}{
    \begin{tikzpicture}
        \startnewset<L>[$\begin{aligned}M_J^{2T}\text{$=$}\sum\limits_{me}L_{me}^{J}t_m^{e}\end{aligned}$]{\diagram{   
            \OPo*[r](0,0)<L>
            \LOOPCWd(L)<a><i>
            \OPt*{1}(i-2)<t1>
        }}
        \nextterm[][$\begin{aligned}M_{aiJ}^{3T}\text{$=$}\sum\limits_fL_{af}^Jt_i^{f}\end{aligned}$]{\diagram{
            \OPo*[r](0,0)<L>
            \LINEuul(L)<e>[left]<a>
            \OPt*{1}(e-1)<t1>
            \LINEINul(t1)<i>
        }}
        \nextterm[][$\begin{aligned}M_{aiJ}^{2TT}\text{$=$}\sum\limits_{em}L_{me}^{J}t_i^{e}t_m^{a}\end{aligned}$]{\diagram{
            \OPo*[r](0,0)<L>
            \LINEVdl(L)<e><m>
            \OPt*{1}(e-1)<ta>
            \LINEINur(ta)<i>[shorten start]
            \OPt*{1}(m-2)<tb>
            \LINEOUTul(tb)<a>
        }}
        \gotonextline[][0][$\begin{aligned}M_{aiJ}^{1T}\text{$=$}\sum_{m} L_{mi}^{J}t_m^{a}\end{aligned}$]{\diagram{
            \OPo*[r](0,0)<L>
            \LINEddl(L)<i>[right]<m>
            \OPt*{1}(m-2)<t1>
            \LINEOUTul(t1)<a>
        }}
        \nextterm[][$\begin{aligned}M_{aiJ}^{2T}\text{$=$}\sum\limits_{me}L_{me}^{J}t_{im}^{ae}\end{aligned}$]{\diagram{
            \OPt*{2}(0,0)<t2>
            \LOOPCCWu(t2-1)<e><m>
            \OPo*[r](e-2)<f>
            \LINEVur(t2-0)[left]<i><a>
        }}
        \nextterm[][$\begin{aligned}\sum\limits_mL_{mb}^Jt_m^{a}\end{aligned}$]{\diagram{
            \OPo*[r](0,0)<L>
            \LINEVdl(L)<b><m>
            \OPt*{1}(m-2)<t1>
            \LINEOUTul(t1)<a>
        }}
        \gotonextline[][0][$\begin{aligned}\sum\limits_eL_{ie}^Jt_j^{e}\end{aligned}$]{\diagram{
            \OPo*[r](0,0)<L>
            \LINEVdr(L)<e><i>
            \OPt*{1}(e-1)<t1>
            \LINEINul(t1)[left]<j>
        }}
    \end{tikzpicture}
    }
\end{figure}

%%%%%%%%%%%%%%%%%%%%%%%%%%%%%%%%%%%
\section{Working equations for EOM-EA and IP}
%%%%%%%%%%%%%%%%%%%%%%%%%%%%%%%%%%%

%
For EOM-EA ($\hat{R}=\hat{R}^{\text{EA}}$) the amplitudes equate to
\begin{equation}
    \begin{split}
        E_{\text{exc.}}r^{a}&=\sum_{e}F_{ae}r^{e}+\sum_{me}F_{me}r_{m}^{ae}+\sum_{Je} M_{ae}^J M_{eJ}^{2R*}\\
    \end{split}
\end{equation}
and
\begin{equation}
    \begin{split}
     E_{\text{exc.}} r_{j}^{ab} & = W_{abj}^{(r)}+\mathcal{P}_{ab}^{-}\Biggl\{\sum_{me}t_{jm}^{ae}\sum_{J}M_{be}^{J}M_{mJ}^{2R*}+\sum_{e}F_{be}r_{j}^{ae}\\
    &+\sum_{J}M_{bj}^J\Biggl(M_{aJ}^{3R*} -M_{aJ}^{2RT*} +M_{aJ}^{2R*}\Biggr)-\sum_{me}r_{m}^{ae} W_{mbej}^{(2t)}\Biggr\}\\
    & -\sum_{m}F_{mj}r_{m}^{ab}+\sum_{m}t_{jm}^{ab}\Biggl(\sum_{e}F_{me}r^{e}+\sum_{Je}L_{me}^{J}M_{eJ}^{2R*}\Biggr)\\
    &-{\frac{1}{2}}\sum_{mn}t_{mn}^{ab}\Biggl(-\mathcal{P}_{mn}^{-}\sum_JM_{nj}^JM_{mJ}^{2R*}+W_{mnj}^{(r)}\Biggr)\;.
    \end{split}
\end{equation}
In analogy to EOM-EE the intermediate of Ref.~\onlinecite{Krylov.2013} $I^{(3i)}_{iab}$ is redefined and added to $I^{(1i)}_{iab}$ resulting in 
\begin{equation}
    {-\sum_{me}r_{m}^{ae}W_{mbej}^{(2t)} =} -\sum_{me}r_{m}^{ae}\Biggl(\sum_JM_{mj}^{J}M_{eb}^{J*} + \sum_{J}W_{mbej}^{(1t)}\Biggr)\;.
\end{equation}
Thus, the scaling is overall $\mathcal{O}(N^5)$ for CD-EOM-EA-CCSD and equates to $V^4N_{\text{CH}}+OV^4+2O^2V^3+2O^2V^2$ which is to be compared to $OV^4+O^2V^3$ for the canonical implementation. As for EOM-EE from the overhead due to recalculation more flops are required but overall memory requirements decrease.
The working equations for EOM-IP ($\hat{R}=\hat{R}^{\text{IP}}$) are
\begin{equation}
    \begin{split}
         E_{\text{exc.}}r_{i}&=-\sum_{m}F_{mi}r_m+\sum_{me}F_{me}r_{mi}^{e}-\sum_{Jm}M_{mi}^{J}M_{mJ}^{2R*}
    \end{split}
\end{equation}
and
\begin{equation}
    \begin{split}
         E_{\text{exc.}} r_{ji}^{b} & =-\sum_{m} W^{(t)}_{mbij}r_m + \mathcal{P}_{ij}^{-} \Biggl[ \sum_m F_{mi} r_{mj}^{b}+\sum_{J}M_{bj}^{J}M_{iJ}^{2R*}\\
        &+ \sum_{me}r_{mj}^{e}\left(  \sum_{J}W_{mbei}^{(2t)} \right)\Biggr]\\
        &+ \sum_{e}F_{be}r_{ji}^{e}+ \frac{1}{2}\sum_{mn}W^{(t)}_{mnij}r_{nm}^{b}-\sum_{e}t_{ij}^{eb}\sum_{Jm}L_{me}^{J}M_{mJ}^{2R*}\;.
    \end{split}
\end{equation}
Again in contrast to Ref.~\onlinecite{Krylov.2013} $I^{(2i)}_{ija}$ is redefined and added to $I^{(1i)}_{ija}$ which results in 
\begin{equation}
   { \sum_{me}r_{mj}^{e} W_{mbei}^{(2t)}= }  \sum_{me}r_{mj}^{e}\left( \sum_JM_{mi}^{J}M_{eb}^{J*}  + \sum_{J}W_{mbei}^{(1t)}\right)\;.
\end{equation}
For the cost of saving an additional $VVOO$ type intermediate $2O^2V^2N_{\text{CH}}+O^3V^2+O^2V^3$ flops are being saved per iteration. Thus, for EOM-IP the scaling of our CD implementation is identical to the canonical one. Both scale as $O^3V^2+O^4V$. 
\begin{table}[tb]
\centering
\caption{CCSD-EOM-EA intermediates}
\label{tab:EOMEA_ccsd_intermediates}
\begin{ruledtabular}
\begin{tabular*}{0.48\textwidth}{@{\extracolsep{\fill}}lclc}
\multicolumn{4}{c}{EOM-EA}  \\
\midrule 
\multicolumn{3}{c}{Equation} & Scaling \\
\midrule
  $W_{ijk}^{(r)}$&=&$-\frac{1}{2}\sum\limits_{ef}\left(\mathcal{P}_{ef}^{-}\sum\limits_{J}L_{ie}^{J}L_{fj}^{J*}\right)r_{k}^{ef}$&$V^2O^3$\\
  $W_{abj}^{(r)}$&=&$-\frac{1}{2}\sum\limits_{ef}\left(\mathcal{P}_{ab}^{-}\sum\limits_{J}M_{af}^{J}M_{eb}^{J*}\right)r_{j}^{ef}$& $V^4O$\\
  $W_{abj}^{(l)}$&=&$-\frac{1}{2}\sum\limits_{ef}\left(\mathcal{P}_{ab}^{-}\sum\limits_{J}M_{fa}^{J}M_{be}^{J*}\right)l_{ef}^{j}$&$V^4O$\\
\end{tabular*}%
\end{ruledtabular}
\end{table}
\begin{table}[tb]
\centering
\caption{$r$ and $l$-transformed Cholesky vectors}
\label{tab:r_transformed_cholvec_IPEA}
\begin{ruledtabular}
\begin{tabular*}{0.48\textwidth}{@{\extracolsep{\fill}}lcllcl}
 \multicolumn{6}{c}{EOM-EA}  \\
 \midrule 
 $M_{aJ}^{2R*}$&=&$\sum\limits_{me}L_{em}^{J*}r_{m}^{ae}$ & $M_{aJ}^{2L*}$&=&$\sum\limits_{me}{ M}_{em}^{J*}l^{m}_{ae}$\\
 $M_{iJ}^{2R*}$&=& $\sum\limits_eL_{ei}^{J*}r^{e}$ & &&\\ 
 $M_{aJ}^{3R*}$&=&$\sum\limits_eL_{ea}^{J*}r^{e}$ & $M_{aJ}^{3L*}$&=&$\sum\limits_e{ M}_{ae}^{J*}l_{e}$ \\
 $M_{aJ}^{2RT*}$&=&$\sum\limits_{me}L_{em}^{J*}r^{e}t_m^{a}$ & &&\\
 \midrule  
 \multicolumn{6}{c}{EOM-IP}  \\
 \midrule 
  $M_{iJ}^{2R*}$&=&$\sum\limits_{me}L_{em}^{J*}r_{mi}^{e}$ & $M_{iJ}^{2L*}$&=&$\sum\limits_{me}{ M}_{em}^{J*}{ l_{e}^{im}}$ \\
   && & $M_{iJ}^{1L*}$&=&$\sum\limits_m{{ M}_{mi}^{J*}l^m}$\\
\end{tabular*}%
\end{ruledtabular}
\end{table}
In the same manner the $l_1$ and $l_2$ amplitudes for EOM-EA are given as
\begin{equation}
    \begin{split}
         E_{\mathrm{exc}}l_{a}     =  & \sum_e F_{ea} l_{e} - \sum_{m} F_{ma} \tilde{l}_{m}+\sum_{Je} M_{ea}^J M_{eJ}^{2L*} \\
                       & + \sum_{Jn} L_{na}^{J} \Biggl(  \frac{1}{2} \sum_{mo} M_{mo}^{J*} \sum_{ef} l_{ef}^{m} t_{no}^{ef}  \\
                       & - \sum_{ge} M_{ge}^{J*} \sum_{mf} t_{nm}^{gf} l_{ef}^{m}\Biggr)
    \end{split}
\end{equation}
and 
\begin{equation}
    \begin{split}
         E_{\mathrm{exc}}l^{j}_{ab}   = & W_{abj}^{(l)}  -\mathcal{P}_{ab}^{-} F_{ja}l_{b} + \mathcal{P}_{ab}^{-} \sum_{e} F_{eb} l_{ae}^{j} - \sum_{m} F_{jm} l_{ab}^{m} \\
                         & { - \mathcal{P}_{ab}^{-} \sum_{me} W_{jebm}^{(2t)} l_{ae}^{m} }+\frac{1}{4}\sum_{mn}\left(\mathcal{P}_{mn}^{-}\sum_{J}L_{ma}^{J}L_{bn}^{J*}\right)\sum_{ef}t_{mn}^{ef}l_{ef}^{j} \\
                         & - \mathcal{P}_{ab}^{-}\sum_J L_{ja}^{J} \Biggl( M_{bJ}^{3L*} + M_{bJ}^{2L*} - \sum_{m} L_{bm}^{J*} \tilde{l}_{m} \Biggr)
    \end{split}
\end{equation}
EOM-IP yields the amplitude equations
\begin{equation}
    \begin{split}
         E_{\mathrm{exc}}l^{i}     =  &- \sum_m F_{im}l^{m}  + \frac{1}{2}\sum_{mne} W_{ienm}^{(t)} l_{e}^{mn}  \\
    \end{split}
\end{equation}
and 
\begin{equation}
    \begin{split}
         E_{\mathrm{exc}}l^{ij}_{b}   = & {\sum_{e}F_{eb} l_{e}^{ij} } -\mathcal{P}_{ij}^{-} F_{ib}l^{j} + \mathcal{P}_{ij}^{-} \sum_{m} F_{im} l_{b}^{jm} \\
                         & + \frac{1}{2} \sum_{mn}W_{ijmn}^{(t)}l_{b}^{mn} -\mathcal{P}_{ij}^{-}\sum_{me} W_{jebm}^{(2t)} l_{e}^{im}\\
                         &  -\mathcal{P}_{ij}^{-} \sum_J L_{ib}^{J} \Biggl( - M_{jJ}^{1L*} + M_{jJ}^{2L*}     - \sum_{e} L_{ej}^{J*} \tilde{l}_{e} \Biggr)\;,\\
    \end{split}
\end{equation}
respectively.

%%%END OF MAIN TEXT%%%


%\balance


%%%REFERENCES%%%
%\bibliography{Literatur} %You need to replace "rsc" on this line with the name of your .bib file
%\bibliographystyle{rsc} %the RSC's .bst file

\providecommand{\noopsort}[1]{}\providecommand{\singleletter}[1]{#1}%
\providecommand*{\mcitethebibliography}{\thebibliography}
\csname @ifundefined\endcsname{endmcitethebibliography}
{\let\endmcitethebibliography\endthebibliography}{}

\end{document}